\documentclass[prb,twocolumn,superscriptaddress,showpacs,floatfix]{revtex4}
\usepackage{amsmath,amssymb,bm,bbm} 
\usepackage{graphicx}  
\usepackage{wasysym}
\usepackage{color} 
\usepackage{relsize} 
\usepackage[utf8]{inputenc}
\usepackage{bbm}
\usepackage{mathrsfs}
\usepackage{epsfig,psfrag}
\usepackage[final]{pdfpages}
\usepackage{pstool}
\usepackage{anyfontsize}
\usepackage{rotating}
\usepackage{longtable}
\usepackage{braket}
\def\be{\begin{equation}}
\def\ee{\end{equation}}
\def\bea{\begin{eqnarray}}
\def\eea{\end{eqnarray}}

\begin{document}

\title{Magnetic field induced topological nodal-lines in triplet excitations of
 frustrated antiferromagnet CaV$_4$O$_9$}
\author{Moumita Deb}\email{moumitadeb44@gmail.com}
\author{Asim Kumar Ghosh}
 \email{asimkumar96@yahoo.com}
\affiliation {Department of Physics, Jadavpur University, 
188 Raja Subodh Chandra Mallik Road, Kolkata 700032, India}
\begin{abstract}
  Magnetic field induced multiple non-Dirac nodal-lines are found
to emerge in the triplet dispersion bands of
a frustrated spin-1/2 antiferromagnetic model on the CaVO lattice.
Plaquette and bond operator formalisms have been employed to obtain
the triplet plaquette and bond excitations for two different 
parameter regimes in the presence of nearest and next-nearest-neighbor
Heisenberg interactions based on the plaquette-resonating-valence-bond
and dimerized ground states, respectively. 
In the absence of magnetic field a pair of six-fold degenerate nodal-loops
with distinct topological feature is noted in the plaquette excitations.
They are found to split into three pairs of two-fold degenerate
nodal-loops in the presence of magnetic field. 
In the other parameter regime, system hosts two-fold degenerate
multiple nodal-lines with a
variety of shapes in the triplon dispersion bands in the
presence of magnetic field. 
Ground state energy and spin gap have been determined
additionally for the two regimes.
Those nodal-lines are expected to be observed in the 
inelastic neutron scattering experiment on
the frustrated antiferromagnet, CaV$_4$O$_9$. 
\end{abstract}
\maketitle
\section{INTRODUCTION}
Studies on topological states of matter are continuing 
with great interest in the recent times 
for the understanding of their various symmetry
protected properties. Topological matter is broadly
classified as topological insulator (TI) and topological semimetal (TSM)
besides the topological superconductor. 
In contrast to TI, TSM has semimetallic bulk state in addition to
metallic surface states for both \cite{Vishwanath,Ali,Fang}.
Three types of TSM are there, Dirac (DSM),
Weyl (WSM) and nodal-line semimetals (NLS). 
DSM and WSM may emerge when band touching occurs
at distinct points on the Brillouin zone (BZ), while
band touching over lines gives rise to NLS. Band touching
implies the degeneracy of the bands which can be studied
in terms of symmetry of the Hamiltonian. 
NLS can be classified into two types: Dirac NLS (DNLS) and Weyl
NLS (WNLS). DNLS is protected by the simultaneous presence of
space inversion (${\cal P}$) and time-reversal (${\cal T}$) 
symmetries which lead to the four-fold degenerate nodal line
for this case. On the other hand, WNLS appear if the system
breaks either ${\cal P}$
or ${\cal T}$ symmetry leading to a pair of two-fold degenerate nodal lines.
In addition, WSM and WNLS could emerge only in
odd spatial dimension \cite{Vishwanath}. 
On further development, DNLS nowadays are classified in terms of
topological protection by separately  
(i) combined ${\cal PT}$, (ii) mirror 
and (iii) non-symmorphic symmetries \cite{Ali}. 
A large number of real materials have been characterized in terms of those
classification norms assigned for TSMs. 
 
Search of topological nodal-line in the magnetic 
excitation modes has been started in the more recent time.
It begins with the finding of a solitary nodal-ring in the magnon bands 
of an antiferromagnetic (AFM)
Heisenberg model on a cubic lattice in the presence of
Dzyloshinskii-Moriya interaction (DMI) where
${\cal P}$ symmetry is broken \cite{FangPRL}. 
Four-fold degeneracy of the nodal-ring in this particular case 
attributes to the fact that AFM ground state constitutes
the bipartite lattice of two oppositely oriented 
spin sublattices. Doubly degenerate nodal-line is
found in magnon dispersion of a ${\cal PT}$ invariant
ferromagnetic (FM) Heisenberg model on 
pyrochlore lattice \cite{Mertig}.
Four-fold degenerate Dirac nodal-loops have been
noted in the AFM magnon dispersions of a ${\cal PT}$ symmetric
Heisenberg model on two-dimensional (2D) square-octagon lattice
based on the superconducting materials,
AFe$_{1.6+x}$Se$_2$ (A$=$K, Rb, Cs) \cite{Owerre}.
For all those models ground states have long-range spin order.  
In another attempt, a pair of six-fold degenerate nodal-rings
have been obtained in the triplet six-spin plaquette excitations
of a frustrated spin-1/2 AFM Heisenberg model on the honeycomb lattice, when 
the ground state lies in a spin-disordered
plaquette-valence-bond-solid (PVBS) phase \cite{Moumita1}.
Recently, U(1)-symmetry protected nodal-loops of triplons are noted
in the AFM Heisenberg model on the Shastry-Sutherland lattice \cite{Dhiman}. 
However, no topological nodal-line is observed in the magnetic excitations
of real materials till now. So, the search of topological
nodal-lines by formulating theoretical models based on real
materials continues. 

\begin{figure*}[t]
    \includegraphics [width=500pt]{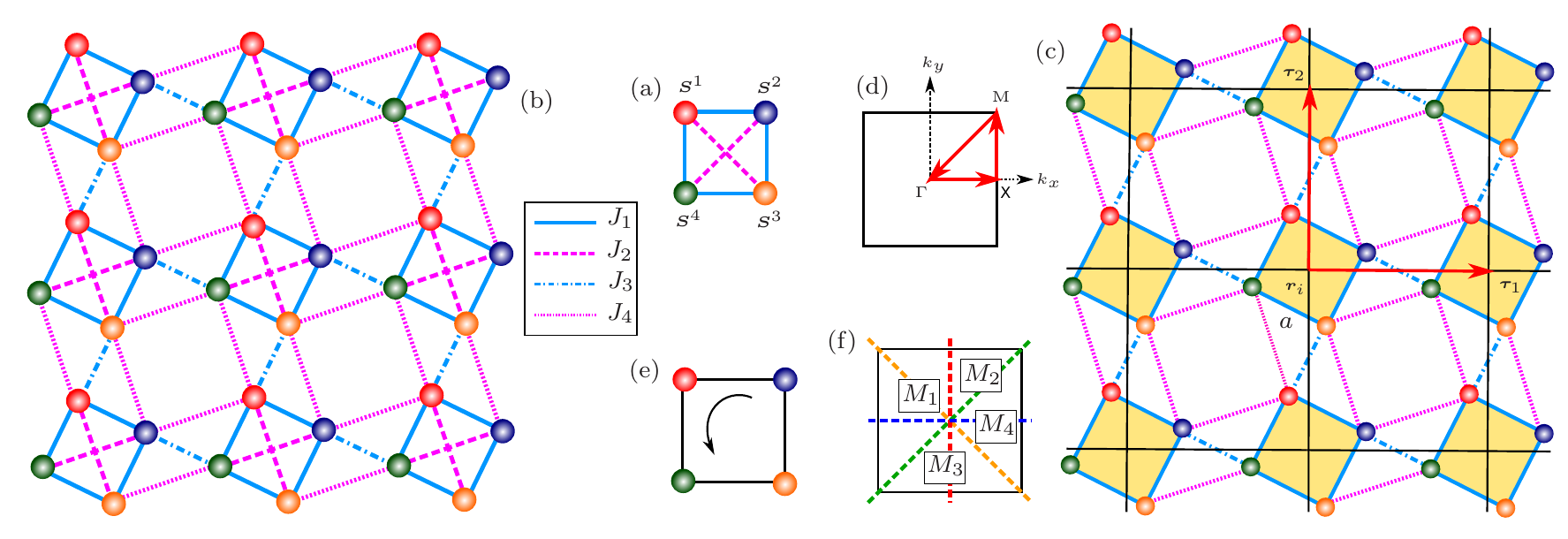}
\caption{(a) Geometrical view of the square plaquette,
  (b) $J_1$-$J_2$-$J_3$-$J_4$ AFM Heisenberg model
on the CaVO  lattice,
(c) effective square lattice constituted by the smallest square plaquettes,
(d) BZ of the square lattice as shown in (c), 
whose corners are defined by $\Gamma=(0,0)$, X$=(\frac{\pi}{\sqrt{5}\,a},0)$
and M$=(\frac{\pi}{\sqrt{5}\,a},\frac{\pi}{\sqrt{5}\,a})$,
(e) system remains invariant under rotation by $\pi/2$ ,
(f) mirror planes are shown by four different dashed lines.}
\label{lattice}
\end{figure*}

CaV$_4$O$_9$ is the first compound with spin gap whose magnetic
property has been explained in terms of the 2D spin-1/2
frustrated AFM Heisenberg Hamiltonian \cite{Taniguchi}.
Spin-1/2 V$^{4+}$ ions in CaV$_4$O$_9$ constitute
a definite form of 1/5-depleted square lattice
which is also known as CaVO lattice. 
This particular non-Bravais lattice can be derived from the square lattice
by removing one-fifth number of its total sites in a particular manner
such that it can be decomposed into
four square sublattices as shown by spheres with four different colors in
Fig \ref{lattice} (b).
As a result, CaVO lattice may be imagined as composed
of identical square-plaquettes,
where each plaquette contains four different sites on its vertices
with one site from each of four sublattices.
One such plaquette is shown in Fig \ref{lattice} (a).
Value of spin gap of this compound has been estimated
before by using a number of theoretical techniques including
a four-spin plaquette operator theory (POT) on CaVO lattice
\cite{Katoh,Ueda,Troyar,Starykh,Takano,Mila,Gelfand,White,Kodama,Ghosh}.
Here spin gap corresponds to the
minimum value of triplet four-spin plaquette excitation energy 
with respect to singlet plaquette-resonating-valence-bond (PRVB)
ground state led by the presence of stronger
intra-plaquette AFM interactions. 
POT on CaVO lattice has been formulated before
by considering intra- and inter-plaquette nearest-neighbor (NN) and
next-nearest-neighbor (NNN) AFM exchange interactions
based on the two-state space constituted by the lowest
singlet and triplet states
of a single plaquette \cite{Starykh}. 

In this investigation, POT has been developed
on an expanded basis space spanned by two singlets
and three triplet plaquette states and when the
system is studied in terms of
weakly interacting plaquettes by means of
stronger NN intra-plaquette terms in the Hamiltonian. 
In another case, bond operator theory (BOT) has been formulated
for the system of weakly interacting bonds 
when the NN inter-plaquette terms in the Hamiltonian are dominant.
Obviously, POT and BOT are based on the two different
ground states, known as PRVB and dimerized states. 
Emergence of a variety of nodal-lines is found in the triplet dispersion
bands both for plaquette and bond excitations obtained in
POT and BOT, respectively, for the frustrated spin-1/2 AFM
Heisenberg model on the CaVO lattice.

A pair of six-fold degenerate nodal-loops of different topological
features emerges in the plaquette excitations,
each of which is found to decouple into three two-fold degenerate loops
in the presence of an external magnetic field. This implies the fact
that every triplet dispersion branch was triply degenerate due to SU(2)
invariance and this degeneracy is withdrawn as soon as the magnetic field
is switched on. 
Among the two loops, one is circular and it is found to appear
at a definite energy,
while the second one is a square and it is spanned
over a finite energy width.
One of the decoupled loops for every case is
found to appear at the same energy values, 
while the remaining two are found to shift their positions towards
higher energies with the increase of magnetic field.
But their overall features remain unaltered 
under the variation of magnetic field. Thus these nodal-lines
are protected by the ${\cal P}$ and U(1) symmetries,
since the ${\cal T}$ symmetry is broken by the magnetic field. 
In case of BOT, system is found to host magnetic field induced
multiple nodal-lines of various shapes within the
two lower bands of triplon excitations,
with the variations of exchange parameters.
None of the nodal-lines are four-fold
degenerate. So, the system is found to host several
non-Dirac nodal-lines, and they seem to be
detectable in the inelastic neutron scattering experiment on the
frustrated antiferromagnetic compound, CaV$_4$O$_9$,
under magnetic field. 

The article is arranged in the following manner.
Section \ref{POT} contains the details of POT.  Properties of
nodal-lines have been described and effect of magnetic field
has been investigated. Similarly, BOT has been described 
in the section \ref{BOT}.
Emergence of nodal-lines under the magnetic field has been
studied. Summary of the results
are presented in the last section (\ref{Discussion}).  
 \section{Four-spin plaquette operator theory}
 \label{POT}
In order to develop the 
four-spin POT applicable for the AFM model on CaVO  lattice,
spin-1/2 operators on the four different vertices of
the square plaquette have been expressed in terms of
the plaquette operators. Thus, 
the $\alpha$-th components of the spin operators at the
$n$-th vertex, $S^n_\alpha$, $n=1,2,3,4$, $\alpha = x, y,z$,
have been expressed in terms of the basis
states spanned by the complete set of eigenstates of a single
frustrated square plaquette $\ket{\eta}$ and  $\ket{\xi}$ 
by following the formula
$S^n_\alpha=\langle \eta|S^n_\alpha|\xi\rangle\ket{\eta}\bra{\xi}$,
where summation over repeated indices is assumed \cite{Sachdev2}.
For this purpose, expressions of the eigenstates 
of a single frustrated square plaquette are obtained. 
\subsection{Single square plaquette}
\begin{figure}[h]
\begin{center}
 \includegraphics[width=230pt]{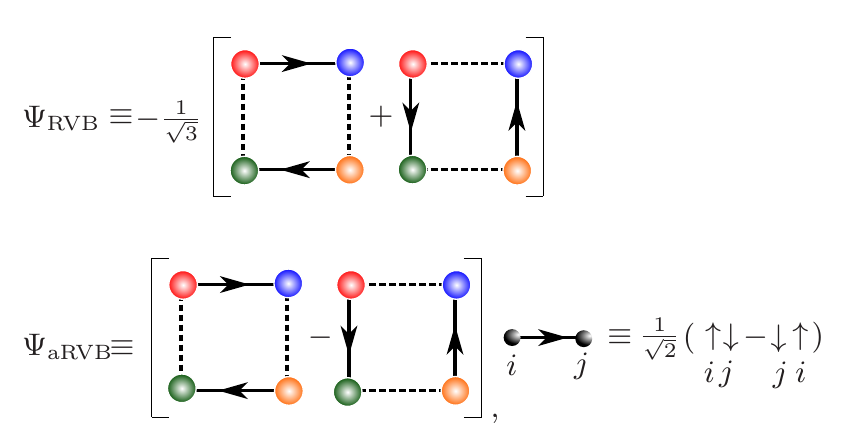}
 \caption{Schematic representation of $\Psi_{\rm RVB}$ and $\Psi_{\rm aRVB}$.
   Arrow on the NN bonds indicates the spin ordering
 in singlet dimers.}
 \label{RVBs}
\end{center}
\end{figure}

The spin-1/2 AFM Heisenberg Hamiltonian on a single square
plaquette is defined by
\begin{eqnarray}
 H_{\text{\Square}}=\!\sum\limits_{n=1}^{4}\left(J_1\, \boldsymbol{S}^n\cdot\boldsymbol{S}^{n+1}
\!+\!J_2\, \boldsymbol{S}^n\cdot\boldsymbol{S}^{n+2}\right),\,
 \boldsymbol{S}^{n+4}\!=\!\boldsymbol{S}^n,
 \label{ham}
\end{eqnarray}
where $\boldsymbol{S}^n$ is the spin-1/2 operator at the position $n$. 
$J_1$ and $J_2$ are the intra-plaquette NN and NNN exchange
interaction strengths, respectively.
A pictorial view of this spin model is shown in Fig \ref{lattice} (a). 
Non-zero values of $J_2$ invokes
frustration in this model since the NNN bonds now 
are not energetically favorable in order to the minimized value of 
classical ground state energy with respect to the NN bonds. 
The Hilbert space of this Hamiltonian 
comprises of 16 states with 
two singlets ($S_{\rm T}=0$), three triplets ($S_{\rm T}=1$) and one 
quintet ($S_{\rm T}=2$), where
$\boldsymbol{S}_{\rm T}=\sum_{n=1}^{4}\boldsymbol{S}^n$ is
the total spin of the plaquette. Two singlets are specified by
$\ket{s_1}$ and $\ket{s_2}$, which can be expressed as
linear combinations of two different plaquette states, where
each plaquette state is composed of two singlet dimers.
In analogy with the conventional pair of bonding and anti-bonding states,
these singlet pair can be termed as resonating-valence-bond (RVB) and
anti-RVB (aRVB) states. 
Forms of the wavefunctions, $\Psi_{\rm RVB}\;(\ket{s_1})$ and
$\Psi_{\rm aRVB}\;(\ket{s_2})$ are shown in Fig \ref{RVBs}.

\begin{figure}[h]
\begin{center}
  \includegraphics[width=250pt]{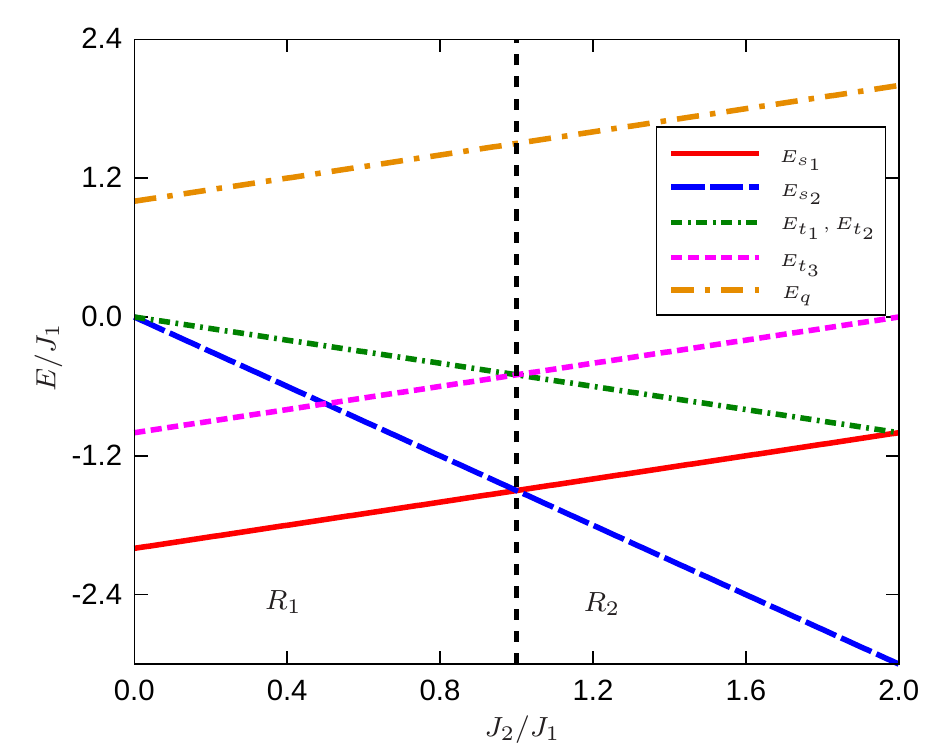}
\end{center}
\caption{Variation of energy eigenvalues of single plaquette
  with $J_2/J_1$. Vertical dashed line separates the two regions,
  R$_1$ and R$_2$.}
  \label{E2D}
 \end{figure}

 The exact analytic expressions of all energy eigenstates 
 along with their symmetries have been presented
 in the Appendix \ref{mean}. 
 Variations of all eigenenergies are shown in Fig \ref{E2D}
 for $0<J_2/J_1<2$, where few crossovers are noted.  
$ E_{s_1}$ ($ E_{s_2}$) is the energy of the singlet state  
$\ket{s_1}$ ($\ket{s_2}$), while  
$ E_{t_1}=E_{t_2}$ and $E_{t_3}$ are the energies 
of the degenerate triplet states $\ket{t_{1,\alpha}}$,  
$\ket{t_{2,\alpha}}$ and $\ket{t_{3,\alpha}}$, respectively.
Quintet has the highest energy, $E_q$, for the entire
region, $0<J_2/J_1<2$, so, it does not cross others. 
Ground state is always a total spin singlet but
not unique in the entire parameter regime, as 
$\ket{s_1}$ and $\ket{s_2}$ are the ground states  
in two separate regions, R$_1$ ($J_2<J_1$) and
R$_2$ ($J_2>J_1$), respectively. 
So, ground state is doubly degenerate at the point, $J_2/J_1=1$.
This indicates the fact that, in contrast to the
property of conventional bonding
and anti-bonding states,
$\Psi_{\rm RVB}$ does not always have energy lower than
$\Psi_{\rm aRVB}$.
Fig \ref{E2D} reveals that two different kinds of spin gap
occurs for a single square plaquette.
One among them can be defined as triplet gap as it
corresponds to singlet-triplet transition  
when $J_2/J_1<1/4$, while singlet gap (singlet-singlet transition)
is found when $J_2/J_1>1/4$.
 
\subsection{Four-spin plaquette operators}
The form of spin-1/2 operators expressed in the truncated basis 
constituted by the lowest singlet and triplet plaquette
states is used before for the estimation of PRVB 
ground state energy ($E_{\rm G}$) and triplet spin gap ($\Delta$) of the
CaVO  lattice \cite{Ueda}. 
The representation of spin-1/2 operators expressed in the basis
constituted by the complete set of plaquette eigenstates
is available \cite{Doretto}. Spin-1/2 operators
in terms of the all singlet and triplet states 
had been employed before for finding the $E_{\rm G}$ and $\Delta$ for the
four-spin PVBS state of the square lattice \cite{Doretto}. 
It is worth mentioning at this time that 
PRVB ground state is unique and it does not break the
symmetry of the CaVO lattice, while
the PVBS state is four-fold degenerate and breaks the
full translation symmetry of the square lattice \cite{Zhitomirsky,Sachdev1}.

Neglecting the contribution of
higher energy quintet states the spin operators, $S^n_\alpha$, are expressed as, 
\begin{equation}
 \begin{aligned}
  S^n_\alpha=&A^n_{a}\left(t^\dagger_{a,\alpha}\,s_{ 1}+
s^\dagger_{1}t_{a,\alpha}\right)+
  B^n_{a}\left(t^\dagger_{ a,\alpha}\,s_2+s^\dagger_2\,t_{a ,\alpha}\right)\\
  &+i\,\epsilon_{\alpha\beta\gamma}\,D^n_{ab}\,t^\dagger_{ a,\beta}\,t_{ b,\gamma}.
  \label{ope}
 \end{aligned}
 \end{equation}
Here, $\epsilon_{\alpha\beta\gamma}$ is the antisymmetric
Levi-Civita symbol with $\epsilon_{xyz}=1$, and $a,\, b =1,2,3$.
The matrix elements, 
$A^n_{a}=\langle s_{1}|S^n_\alpha|t_{a ,\alpha}\rangle$, 
$B^n_{a}=\langle s_{2}|S^n_\alpha|t_{a ,\alpha}\rangle$, and 
$D^n_{ab}=\langle t_{a ,\beta} |S^n_\alpha|t_{b,\gamma}\rangle$,  
are given in the Appendix \ref{mean}.
 The vacuum $\ket{0}$ and five
plaquette operators are defined here which yield the five physical states 
$\ket{s_j}= s^\dagger_j\ket{0},\ket{t_{a,\alpha}}= t^\dagger_{a,\alpha}\ket{0}$,
where $j=1,2$ and $a=1,2,3$. Plaquette operators obey bosonic
commutation relations. 
The completeness relation in the truncated Hilbert space reads as,
\begin{eqnarray}
 \sum_{j=1,2} s^\dagger_j\,s_j  + 
\sum_{a,\alpha} t^\dagger_{a,\alpha}\,t_{a,\alpha}=1. 
 \label{constraint}
 \end{eqnarray}
The Hamiltonian of a single plaquette, Eq \ref{ham}, in the same Hilbert
space assumes the form
 \begin{equation}
 \begin{aligned}
H_{\text{\Square}}&=\sum_{\substack{j}}E_{s_j}s^\dagger_j\,s_j 
+ \sum_{\substack{a,\alpha}} 
E_{t_a}\,t^\dagger_{a,\alpha}\,t_{a,\alpha}.  
 \end{aligned}
 \end{equation}
 
\subsection{Hamiltonian in terms of plaquette operators}
In order to obtain the values of $E_{\rm G}$ and $\Delta$ for the PRVB phase
as well as the triplet dispersion bands for the fully frustrated
spin-1/2 AFM Heisenberg model on the CaVO  lattice,  
the following Hamiltonian has been considered. 
\begin{equation}
 \begin{aligned}
 H\!\!=\!\!&\sum\limits_{i}\!\big[H_{\text{\Square}}(\boldsymbol{r}_i)\!
+\!J_3\!\left(\boldsymbol{S}^2_{\boldsymbol{r}_i}\!\!\cdot\!\boldsymbol{S}^4_{\boldsymbol{r}_i+\boldsymbol{\tau}_1}\!\!
+\!\boldsymbol{S}^1_{\boldsymbol{r}_i}\!\!\cdot\!\boldsymbol{S}^3_{\boldsymbol{r}_i+\boldsymbol{\tau}_2}\right)
 \!\!+\!\!J_4\!\big(\boldsymbol{S}^2_{\boldsymbol{r}_i}\!\!\cdot\!\boldsymbol{S}^1_{\boldsymbol{r}_i+\boldsymbol{\tau}_1}\!\\
&+\!\boldsymbol{S}^3_{\boldsymbol{r}_i}\!\!\cdot\!\boldsymbol{S}^4_{\boldsymbol{r}_i+\boldsymbol{\tau}_1}
+\boldsymbol{S}^2_{\boldsymbol{r}_i}\!\!\cdot\!\boldsymbol{S}^3_{\boldsymbol{r}_i+\boldsymbol{\tau}_2}\!
+\!\boldsymbol{S}^1_{\boldsymbol{r}_i}\!\!\cdot\!\boldsymbol{S}^4_{\boldsymbol{r}_i+\boldsymbol{\tau}_2}\big)\big].
 \label{ham1}
 \end{aligned}
\end{equation}
$J_3$ and $J_4$ are the inter-plaquette NN and NNN exchange
interaction strengths, respectively.
Here, the vector $\boldsymbol{r}_i$ denotes the position of 
the $i$-th plaquette while 
$\boldsymbol{\tau}_1=\sqrt{5}\,a\,\hat{x}$ and
$\boldsymbol{\tau}_2=\sqrt{5}\,a\,\hat{y}$
are the primitive vectors of the effective square lattice
formed by the square plaquettes. 
 Here $a$ is the NN lattice spacing of the original CaVO  
 lattice which has been assumed henceforth unity.
 POT is valid as long as intra-plaquette interactions are
 stronger than respective inter-plaquette interactions, {\em i. e.}, 
 $J_1>J_3$ and  $J_2>J_4$.
 Here the system is studied in a wider parameter regime, $0<J_2/J_1<2$,
 in comparison to all the previous studies where  
 this regime was limited to $0<J_2/J_1<1$. 
 However, in this case, system is studied in terms of weakly
 interacting square plaquettes.
 In other case, BOT is valid for the
 opposite limit, $J_1<J_3$, which will be
 described in the next section.  

 The plaquette-operator representation (Eq \ref{ope}) has been substituted
 in Eq \ref{ham1} to express the Hamiltonian in terms of the
 plaquette operators, 
   \begin{equation}
  \begin{aligned}
  {\cal H}\!=\!E_0\!+\!H_{02}\!+\!H_{20}\!+\!H_{30}\!+\!H_{21}\!+\!H_{40}\!+\!H_{22},   
     \end{aligned}
     \label{ham2}
   \end{equation}
where $E_0$ is a constant.  
In $H_{nm}$, $n$ and $m$ indicate the number of triplet and singlet operators,
whose explicit forms are given in the Appendix \ref{mean}. 
The effect of the constraint (Eq \ref{constraint}) has been
taken into account by including 
the following term to the Hamiltonian (Eq \ref{ham2}), 
    \begin{equation}
  \begin{aligned}
 -\mu\sum\limits_{i}\bigg( \sum\limits_{j=1,2} s^\dagger_{j,i}\,s_{j,i}  + 
\sum\limits_{a,\alpha} t^\dagger_{a,i,\alpha}\,t_{a,i,\alpha}-1\bigg), 
     \end{aligned}
 \end{equation}
where $\mu$ is the Lagrange multiplier.
To execute plaquette operator formalism, the lowest-energy singlet
is assumed to be condensed
and has to be replaced by a number in Eq \ref{ham2}.
It is worthy to mention at this point that $\ket{s_1}$ and $\ket{s_2}$ are the
lowest energy singlets in the regions, R$_1$ and R$_2$, respectively. 
The condensation is implemented by making the accompanying substitution, 
$s^\dagger_{j,i}=s_{j,i}=\langle s^\dagger_{j,i}\rangle
=\langle s_{j,i}\rangle=\bar s$ \cite{Sachdev2}.
The lowest singlet in the respective region has been condensed, which
implies that $\ket{s_j}$ is condensed in the region R$_j$,
where $j=1,2$. 
Now the value of the constant, $E_0$ is given by the equation, 
$E_0=N\left[\bar s^2E-\mu\left(\bar s^2-1\right)\right]$, in which 
$E=E_{s_{ 1}}$ ($E_{s_{2}}$) for the region R$_1$ 
(R$_2$) and $N$ is the total 
number of plaquettes in the system.
By assuming the periodic boundary condition, 
Fourier transformations of the operators $t^\dagger_{a,i,\alpha}$ and $s^\dagger_{j,i}$
are obtained as, 
\begin{equation}
\begin{aligned}
  t^\dagger_{a,i,\alpha}&=\frac{1}{\sqrt{N}}\sum\limits_{\bold{k}}
\text{exp}\left(-i\bold{k}\cdot\boldsymbol{R}_i\right)t^\dagger_{a,\bold{k},\alpha},\\
s^\dagger_{j,i}&=\frac{1}{\sqrt{N}}\sum\limits_{\bold{k}}\text{exp}
\left(-i\bold{k}\cdot\boldsymbol{R}_i\right)s^\dagger_{j,\bold{k}}, 
\end{aligned}
\end{equation}
where $j=1,2$ and $a=1,2,3$.
Here the momentum sum runs over the BZ of the square lattice,
which is shown in Fig \ref{lattice} (d).
\subsection{Quadratic approximation}
\begin{figure}[h]
 \begin{center}
     \includegraphics[width=250pt]{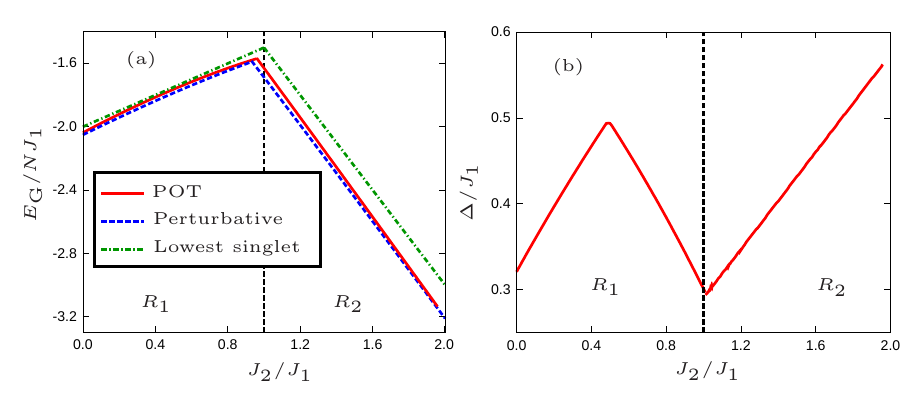}
 \end{center}
 \caption{Variations of ground state energy per plaquette (a) and
   spin gap (b) against
 $J_2/J_1$, when $J_3=J_1/2$ and $J_4=J_2/2$.}
     \label{Epla}
  \end{figure}

Henceforth, the system has been studied in terms of
an effective boson model where
the Hamiltonian keeps only those terms which are
quadratic in bosonic plaquette operators. 
Terms containing more than two
plaquette operators, $H_{30},H_{21},H_{40}$ and $H_{22}$,
therefore, have been neglected. So, the approximated 
Hamiltonian becomes, 
 \begin{equation}
  \begin{aligned}
  {\cal H}_{\rm Q}= E_0+H_{02}+H_{20}.\nonumber
  \end{aligned}
\end{equation}
In the momentum space it becomes
\bea
    {\cal H}_{\rm Q}&=&\sum\limits_{\bold{k}}\left(E_{s_j}-
    \mu\right)s^\dagger_{j,\bold{k}}s_{j,\bold{k}}+
    \sum\limits_{\bold{k}} X^{ab}_{\bold{k}} \;t^\dagger_{a,\bold{k},\alpha}t_{b,\bold{k},\alpha}\nonumber \\
    &+&\frac{Y^{ab}_{\bold{k}}}{2}\!\left(t^\dagger_{a,\bold{k},\alpha}t^\dagger_{b,-\bold{k},\alpha}+
  t_{a,-\bold{k},\alpha}t_{b,\bold{k},\alpha}\!\right), 
   \label{ham3}
 \eea
 with $j=2 \,(1)$ for the region R$_1$ (R$_2$),
 $a,b= 1,2,3$ and $\alpha,\beta,\gamma = x,y,z$.
Expressions of the coefficients $X^{ab}_{\bold{k}}$ and $Y^{ab}_{\bold{k}}$
are given in Appendix \ref{mean}.  
Singlet sector is diagonalized separately
with singlet energies $\Omega_{s_j}=\left( E_{s_j}-\mu\right)$. 
The six-component vector $\Psi^\dagger_{\bold{k},\alpha}=\left
(t^\dagger_{1,\bold{k},\alpha} t^\dagger_{2,\bold{k},\alpha} t^\dagger_{3,\bold{k},\alpha}
t_{1,\bold{- k},\alpha} t_{2,\bold{- k},\alpha} t_{3,\bold{- k},\alpha}\right)$
 is introduced for the diagonalization of the triplet sector.
Eq \ref{ham3} becomes
 \begin{equation}
  \begin{aligned}
 {\cal H}_{\rm Q}=E^\prime_0+\sum\limits_{\bold{k}}\left(E_{s_j}-
    \mu\right)s^\dagger_{j,\bold{k}}s_{j,\bold{k}}+\frac{1}{2}\sum\limits_{\bold{k}}\Psi^\dagger_{\bold{k},\alpha}{H_\bold{k}}\Psi_{\bold{k},\alpha},\nonumber
   \label{ham4}
  \end{aligned}
 \end{equation}
where
 \begin{equation}
 \begin{aligned}
   E^\prime_0=E_0-\frac{3}{2} \sum\limits_{\bold{k}}\; \sum_{\substack{a= 1, 2 , 3 }}   X^{aa}_{\bold{k}}, \; {\rm and}\;
   {H_\bold{k}}=
 \left( 
 { \begin{array}{cc}
  {X_\bold{k}} &  {Y_\bold{k}}  \\
  {Y_\bold{k}} &  {X_\bold{k}} \\
  \end{array}}
 \right).
 \label{hk}
 \end{aligned}
 \end{equation}
 ${X_\bold{k}}$ and ${Y_\bold{k}}$ are two different $3 \times 3$ hermitian
 matrices having elements $X^{ab}_\bold{k}$ and $Y^{ab}_\bold{k}$, respectively.
After diagonalization the Hamiltonian assumes the form
\begin{equation}
  \begin{aligned}
  {\cal H}_{\rm Q}=E_{\rm G}+\sum\limits_{\bold{k}}\left(E_{s_j}-
   \mu\right)s^\dagger_{j,\bold{k}}s_{j,\bold{k}}+
   \frac{1}{2}\sum\limits_{\bold{k}}\Phi^\dagger_{\bold{k},\alpha}
        {H^\prime_\bold{k}}\Phi_{\bold{k},\alpha}, \nonumber
  \end{aligned}
\end{equation}
with the energy of the PRVB (ground) state 
\begin{equation}
  \begin{aligned}
   E_{\rm G}=E_0+\frac{3}{2}\sum\limits_{a,\bold{k}}\left(\Omega_{a,\bold{k}}-X^{aa}_{\bold{k}}\right), 
  \end{aligned}
\end{equation}
and
${H^\prime_\bold{k}}$ = Diag $[h_\bold{k},-h_\bold{k}]_{(2\times 2)}$,  
where
$h_\bold{k}$ = Diag $[\Omega_{1,\bold{k}},\Omega_{2,\bold{k}},
  \Omega_{3,\bold{k}}]_{(3\times 3)}$. 
The eigenvectors $\Phi^\dagger_{\bold{k},\alpha}$ is given by $\Phi^\dagger_{\bold{k},\alpha}=\left (b^\dagger_{1,\bold{k},\alpha} b^\dagger_{2,\bold{k},\alpha} b^\dagger_{3,\bold{k},\alpha}
b_{1,\bold{- k},\alpha} b_{2,\bold{- k},\alpha} b_{3,\bold{- k},\alpha}\right)$.
Relation between the vectors, $\Psi^\dagger_{\bold{k},\alpha}$
and $\Phi^\dagger_{\bold{k},\alpha}$ is established by the transformation,   
\begin{equation}
  \begin{aligned}
   \Phi_{\bold{k},\alpha}= M_{\bold{k}}   \Psi_{\bold{k},\alpha},\quad {\rm where} \quad
    M_\bold{k}=
 \left( 
 { \begin{array}{cc}
  {U^\dagger_\bold{k}} &  {-V^\dagger_\bold{k}} \\
  {-V^\dagger_\bold{k}} &  {U^\dagger_\bold{k}} \\
  \end{array}}
\right). \nonumber
  \end{aligned}
\end{equation}
$ {U^\dagger_\bold{k}}$ and $ {V^\dagger_\bold{k}}$ are two 
$3 \times 3$ hermitian matrices whose elements are
the Bogoliubov coefficients $u^{ab}_\bold{k}$ and $v^{ab}_\bold{k}$, respectively.
The analytic expressions of the triplet excitation energies
$\Omega_{a,\bold{k}}$, along with $u^{ab}_\bold{k}$ and $v^{ab}_\bold{k}$
in terms of the $X^{ab}_\bold{k}$ and $Y^{ab}_\bold{k}$
are available in Appendix \ref{mean}.
The values of $\mu$ and $\bar s^2$ are determined by
minimizing the ground state energy with respect to themselves, {\em i. e.}, 
$\frac{\partial  E_{\rm {G}}}{\partial \bar s^2}=0$
and $\frac{\partial E_{\rm {G}}}{\partial \mu}=0$, which gives 
\begin{equation}
  \begin{aligned}
&\mu=E+\frac{3}{2N}\sum\limits_{a,\bold{k}}
\left[\frac{\partial \Omega_{a,\bold{k}}}{\partial \bar s^2}-\frac{Y^{aa}_{\bold{k}}}{\bar s^2}\right], \\
& \bar s^2=1+\frac{3}{2N}\sum\limits_{a,\bold{k}}
\left[\frac{\partial \Omega_{a,\bold{k}}}{\partial \mu}+1\right].
\label{selfcon}
  \end{aligned}
\end{equation}
Singlet, $\Omega_{s_j}= E_{s_j}-\mu$ and triplet, $\Omega_{a,\bold{k}}$ energies
have been determined after finding the values of $\mu$ and $\bar s^2$
numerically by solving the self-consistent Eq \ref{selfcon}.
In the absence of inter-plaquette interactions,
($J_3=0$, $J_4=0$), $\bar s^2=0$ and $\mu=E$.
In this limit, $E_{\rm {G}}=E_{s_j}$, the value of
ground state energy per plaquette becomes equal to
the energy of corresponding lowest singlet state. 
Variations of ground state energy per plaquette, $E_{\rm G}/NJ_1$ 
with respect to $J_2/J_1$, when $J_3=J_1/2$ and $J_4=J_2/2$ is shown in
Fig \ref{Epla} (a), 
along with the energy of the lowest singlet plaquette state.
$E_{\rm G}/NJ_1$ is always lower than the lowest singlet energy. 
Variation of $E_{\rm G}/NJ_1$ with  $J_2/J_1$ shows resemblance with
that of plaquette singlet energies as shown in Fig \ref{E2D},
in the respective regions. Although, $E_{s_1}$ and $E_{s_2}$
meet at $J_2/J_1=1$ (Fig \ref{E2D}), $E_{\rm G}/NJ_1$s of the two
PRVB states based on the singlets $\ket{s_1}$ and $\ket{s_2}$
for the respective regions, R$_1$ and R$_2$ are found
to meet at $J_2/J_1=0.96$ (Fig \ref{Epla} (a))
due to the inter-plaquette interactions.
This meeting point must vary with the values of $J_3$ and $J_4$. 

The value of $E_{\rm G}$ is further obtained by the
second-order perturbative calculation. In this formulation,
\(H_0=\sum_{i=1}^N H_{\text{\Square}}(\boldsymbol{r}_i)\), has been
treated as the unperturbed Hamiltonian, where $H-H_0$ 
is the perturbation. Here, $H_0$ means the sum of all plaquette Hamiltonians,
while  $H-H_0$ implies the sum of all inter-plaquette interactions.
Obviously, this result is acceptable 
as long as intra-plaquette interaction strengths are
dominant. 
The ground state energy per plaquette for the regions R$_1$ and R$_2$
has been obtained, where, 
  \begin{equation}
  \begin{aligned}
   E^{\rm P}_{\rm G}(\rm R_1)&=\Big(-2J_1+J^\prime_2+6\frac{(\frac{J_3+J^\prime_2}{12})^2}{-4J_1+2J_2}\\
   &-3\frac{(\frac{J_3+J^\prime_2-2J_4}{3})^2}{2J_1}+6\frac{(\frac{J_3+J^\prime_2}{6\sqrt{2}})^2}{-3J_1+J_2}\Big),\\
   E^{\rm P}_{\rm G}(\rm R_2)&=\Big(-3J^\prime_2-3\frac{(\frac{J_3+J^\prime_2}{4})^2}{J_2}\Big),
  \end{aligned}
\end{equation}
with $  J^\prime_2=J_2/2$.
Variation of $E^{\rm P}_{\rm G}/NJ_1$ with 
$J_2/J_1$ has been shown by dashed (blue) line in Fig \ref{Epla} (a), 
which is found to agree with that obtained in POT
shown by red line.

\begin{figure}[h]
\begin{center}
 \includegraphics[width=250pt]{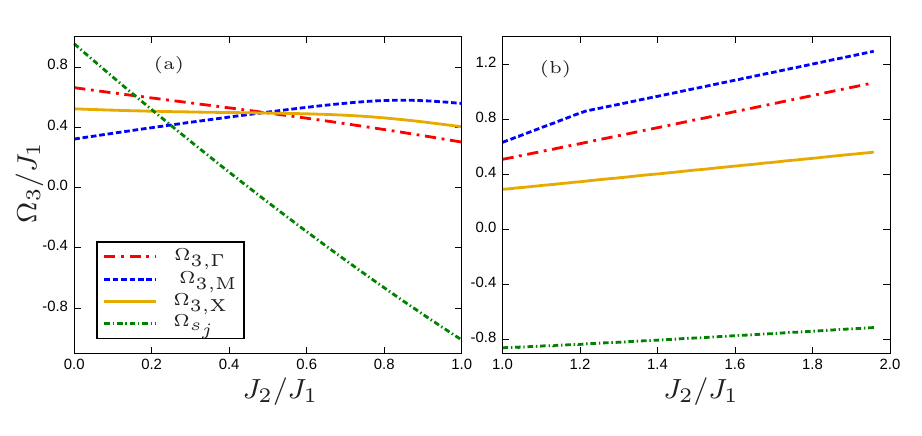}
 \end{center}
   \caption{ Variation of energies of  
 $ \Omega_{3,\Gamma}$, $ \Omega_{3,{\rm M}}$ and $ \Omega_{3,{\rm X}}$ 
measured with respect to $E_{\rm G}$ against $J_2/J_1$ (a) in region R$_1$,
(b) in region R$_2$, when $J_3=J_1/2$ and $J_4=J_2/2$.}
\label{gap}
\end{figure}

Triplet dispersions as shown in the Figs \ref{triplet1}, \ref{triplet2}
reveal that value of $\Omega_{3,\bold{k}}$ is always the lowest.
In addition, its minima
occur at the high-symmetry points, $\Gamma$, M and X in the BZ, 
depending on the values of $J_2/J_1$, which is shown in Fig \ref{gap},
where variations of 
$\Omega_{3,\Gamma}$, $\Omega_{3,{\rm M}}$ and $\Omega_{3,{\rm X}}$ have been
plotted with respect to $J_2/J_1$.
In order to estimate the magnitude of $\Delta$, values of $\Omega_{3,\Gamma}$,
$\Omega_{3,{\rm M}}$ and $\Omega_{3,{\rm X}}$ are measured
with respect to $E_{\rm{G}}$ in Figs \ref{gap} (a) and (b), for the
regions R$_1$ and R$_2$, respectively, where $J_3=J_1/2$ and $J_4=J_2/2$. 
Singlet excitation, $\Omega_{s_2}$ in R$_1$ or $\Omega_{s_1}$ in R$_2$ 
is always dispersionless in the quadratic approximation irrespective
of the values of $J$s.
$\Omega_{3,\Gamma}$, $\Omega_{3,{\rm M}}$ and $\Omega_{3,{\rm X}}$ cross each other
at a single point, $J_2/J_1=1/2$ in R$_1$, as a special case
when $J_3=J_1/2$ and $J_4=J_2/2$. It means that
$\Omega_3$ at the points $\Gamma$, M and X has the same value.  
The plot shows that $\Omega_{3,{\rm M}}$ is the lowest when $J_2/J_1<1/2$
while $\Omega_{3,\Gamma}$ is that when $J_2/J_1>1/2$ in R$_1$.
For different values of  $J_3$ and $J_4$,
$\Omega_{3,\Gamma}$, $\Omega_{3,{\rm M}}$ and $\Omega_{3,{\rm X}}$ cross each other at different points. 
On the other hand, $ \Omega_{3,{\rm X}}$ is always the lowest in R$_2$.
Ultimately, the value of $\Delta$ has been estimated form this comparative study. 
Variations of $\Delta$ 
with respect to $J_2/J_1$, when $J_3=J_1/2$ and $J_4=J_2/2$ is shown in
Fig \ref{Epla} (b).

\begin{figure*}
  \includegraphics[width=500pt]{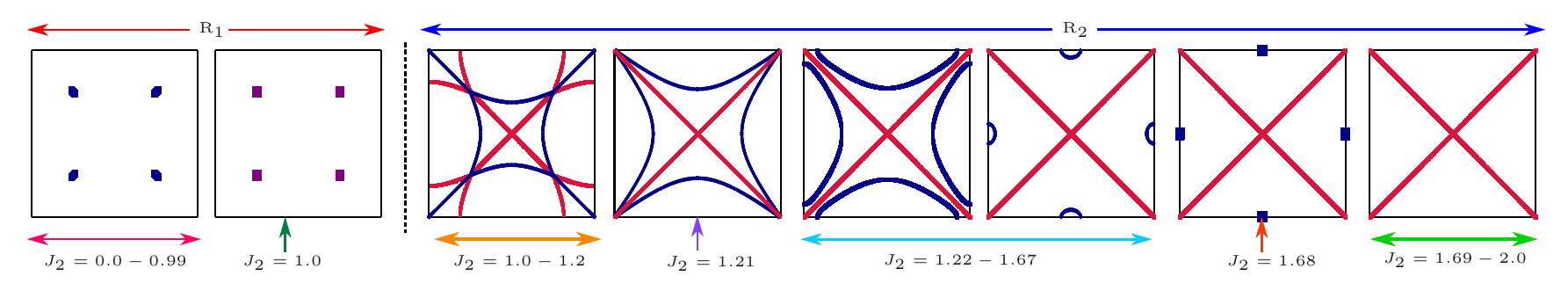}
\caption{Kaleidoscopic views of the emergent topological nodes and
 nodal-lines on the BZ with the change of $J_2/J_1$, covering both
 the regions R$_1$ and R$_2$, when $J_3=J_1/2$ and $J_4=J_2/2$.}
\label{node}
\end{figure*}
\subsection{Triplet plaquette dispersions} 
\begin{figure}[h]
\centering
\includegraphics[width=250pt]{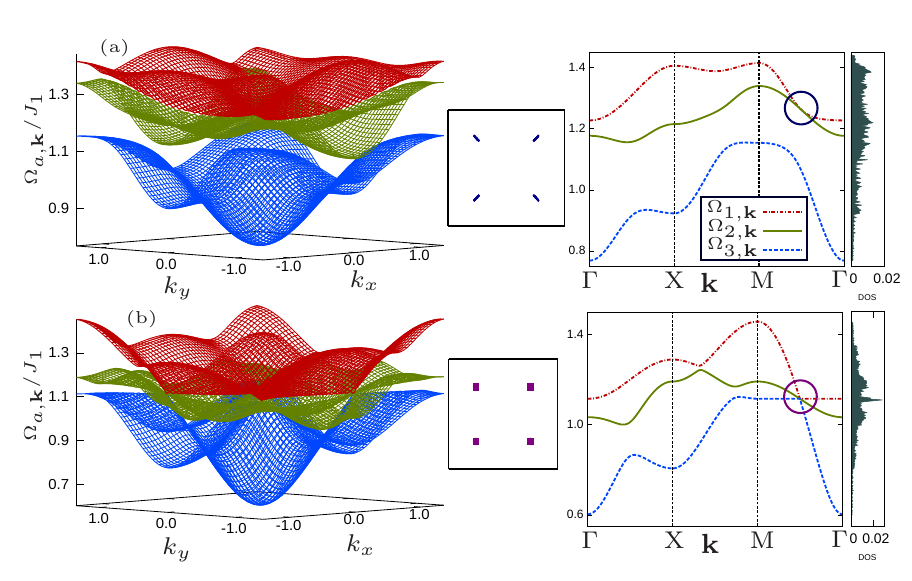}
 \caption{Triplet dispersion bands in R$_1$ when $J_2/J_1=0.8$,
  (a) 1.0, (b), for $J_3=J_1/2$ and $J_4=J_2/2$. }
\label{triplet1}
\end{figure}

\begin{figure}[h]
\centering
\includegraphics[width=250pt]{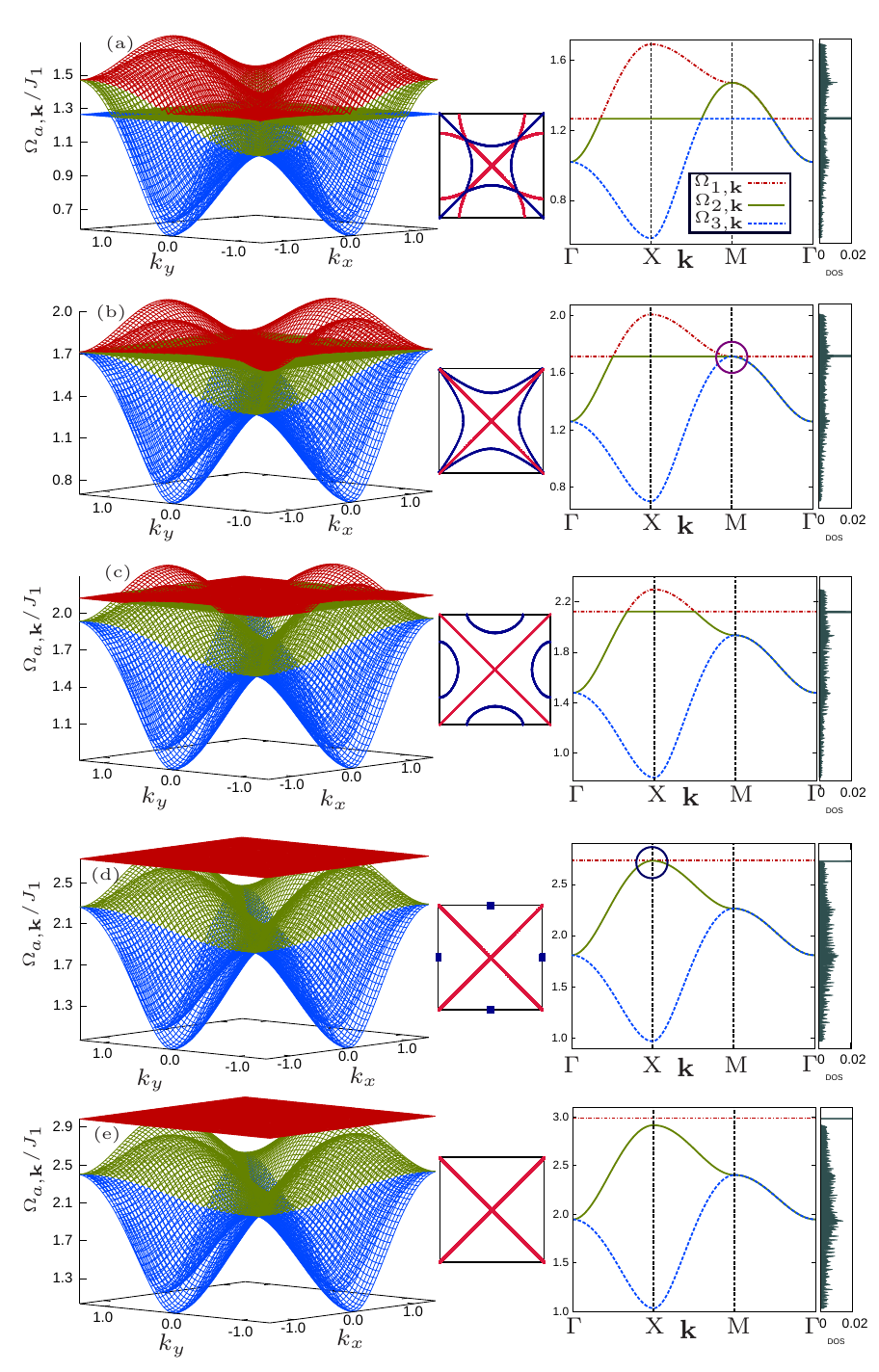}
\caption{Triplet dispersion in R$_2$ with $J_2/J_1$,
  (a) 1.0, (b) 1.21, (c) 1.4, (d) 1.68, (e) 1.8. Here $J_3=J_1/2$ and $J_4=J_2/2$. }
 \label{triplet2}
\end{figure}

Now, emergence and evolution of topological nodes and nodal-lines will be  
described in great detail for the two different regions. 
For this purpose, the triplet dispersion bands,
$\Omega_{a,\bold{k}}/J_1$, $a= 1,2,3$,
have been shown in Figs \ref{triplet1} (a)-(b)
and Figs \ref{triplet2} (a)-(e), for the regions R$_1$
and R$_2$, respectively.
The dispersions are shown in three-dimensional (3D) plots covering the full
BZ, as well as, along the high-symmetry pathway $(\Gamma$,X,M,$\Gamma)$
for every case.  Density of states (DOS) is shown in the
side panel. 
Two types of band-touching points or nodes, 
are found depending on the number of bands at the touching point. 
They are termed as two-band (2BTP) and three-band touching point (3BTP), 
where two and three bands are found to meet, respectively. 
Two kinds of 3BTP are identified owing to their
dissimilar nature of dispersion 
relation around the respective meeting points.
Emergence and evolution of those nodes and nodal-lines 
with the variation of $J_2/J_1$ have been clearly shown 
in Fig \ref{node}. 2BTP and 3BTP appear both in the 
regions R$_1$ and R$_2$, whereas, nodal-line
and flat-band appear only in region R$_2$. 

\begin{figure}[h]
 \includegraphics[width=240pt]{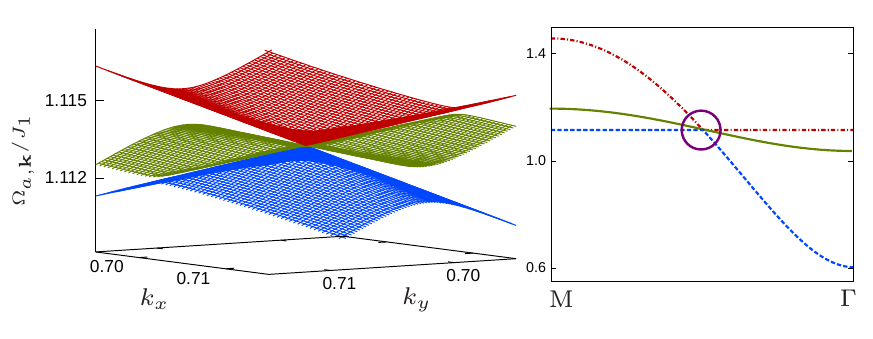}
 \caption{Closer view of triplet dispersions around the
   3BTP for $J_2/J_1=1$ in R$_1$.}
     \label{3BTP}
  \end{figure}  

\begin{figure}[h]
  \includegraphics[width=240pt]{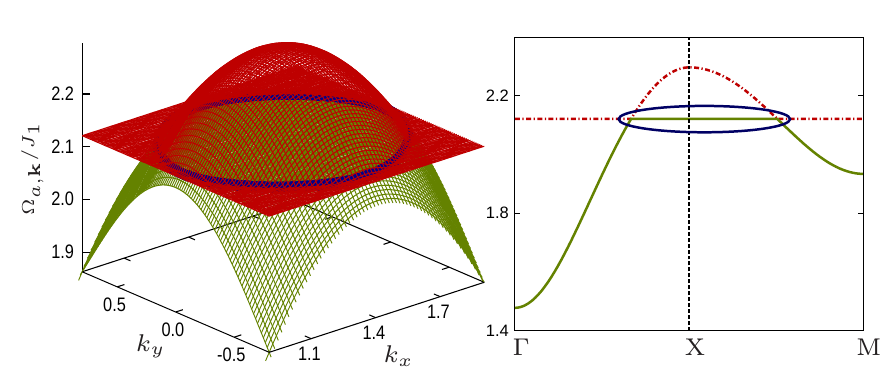}
\caption{Closer view of triplet excitations around the circular 
nodal-line for $J_2/J_1=1.4$.}
     \label{nodal_ring}
  \end{figure} 
\begin{figure}[h]
  \includegraphics[width=240pt]{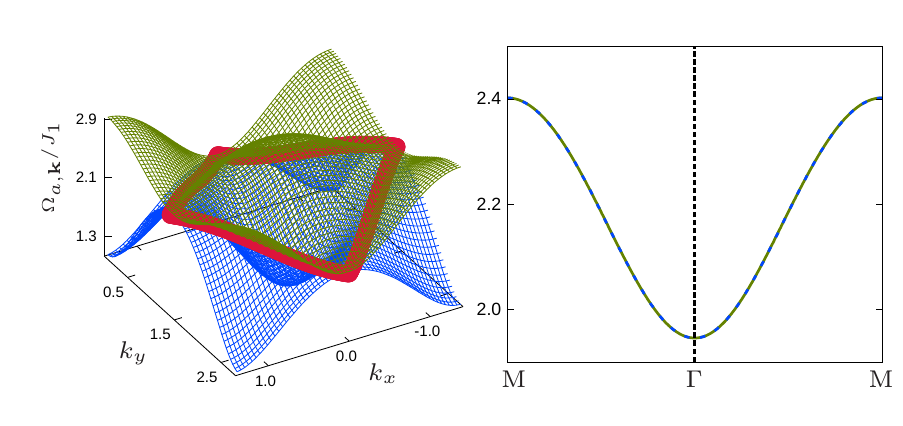}
\caption{Closer view of triplet excitations around the square 
nodal-line for $J_2/J_1=1.8$.}
     \label{nodal_line}
\end{figure}

In R$_1$, a 2BTP in the upper two bands
is noted around the point $(\pi/2\sqrt{5},\pi/2\sqrt{5})$
in the BZ, as long as $0\leq J_2/J_1<1$,
which is shown by blue diamond in Fig \ref{triplet1} (a).
This 2BTP is replaced by a 3BTP, when $J_2/J_1=1$,
as shown by purple square in Fig \ref{triplet1} (b).
They do not therefore coexist.  
All the 2BTP and 3BTP are shown by open circles
along the $(\Gamma$,X,M,$\Gamma)$ pathway. 
Closer view of triplet dispersions around this 3BTP is shown 
in Fig \ref{3BTP}.

The system in region R$_2$ hosts two concentric nodal loops with
different shapes.
They are centered around the X point of the BZ.
Among them one is perfectly square and it forms
between the lower two bands. Area of the
square is exactly equal to the area of the BZ, since it passes through the
high-symmetry points, $\Gamma$, M, centering the X. 
The shape, position and area of this
nodal-loop remain unchanged regardless the values of $J_2/J_1$
for the entire region R$_2$. Thus, it seems that it is additionally
protected by some intrinsic symmetry of the system. 
The other loop is found between the upper two bands and it appears exactly
circular when $J_2/J_1=1.4$. The area of this loop
is always less than that of the BZ. With the increase of 
$J_2/J_1$, radius of this loop decreases and becomes a point
giving rise to a 2BTP, when $J_2/J_1=1.68$.
The bands get separated by leaving a gap with the further increase of 
$J_2/J_1$ beyond the value 1.68.
Thus, this loop is not protected by the intrinsic 
symmetry of the system. 
Fig \ref{nodal_ring} and \ref{nodal_line} show
the magnified views of those nodal loops.
It reveals that circular loop occurs at a definite energy,
while the square loop spans over a energy width. 
The system exhibits a 3BTP at the M point of BZ in this region
which occurs at a definite value, $J_2/J_1=1.21$.
Features of the 3BTPs in R$_1$ and R$_2$ are different.  
Further, the topmost band becomes flat for the regime
$1.68 \leq J_2/J_1 \leq 2$. A sharp peak in the DOS indicates the 
value of energy of this flat band. 
Coexistence of 2BTP and 3BTP with the nodal-loops
is observed in this region. 
Non-vanishing DOS in those band-structure reveals that there is no
true band gap within those triplet dispersion branches. 
All the nodes and nodal-lines are protected by the
${\cal PT}$ and SU(2) symmetries,  
since the Hamiltonian (Eq \ref{ham1}) preserves those symmetries.
Also, all of them are six-fold degenerate, since each
triplet dispersion is three-fold degenerate because of the
SU(2) invariance of the system.

\begin{figure}[h]
\includegraphics[width=260pt]{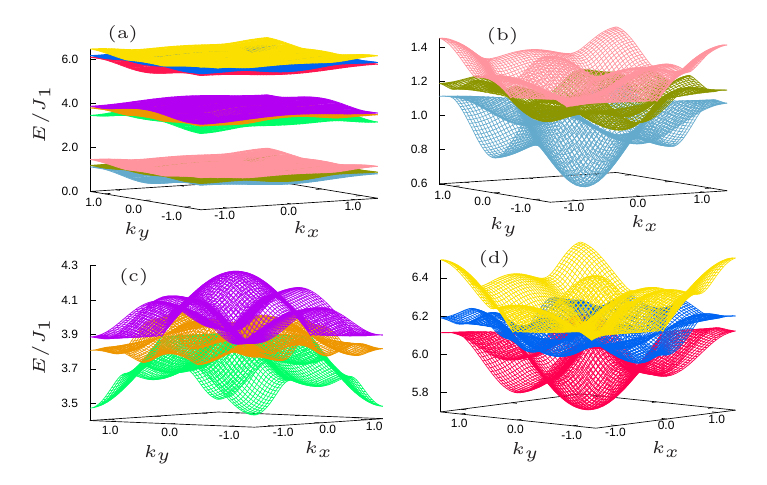}
\caption{(a) Nine triplet dispersion bands for $h_z/J_1=3.0$ and 
$J_2/J_1=1$ in R$_1$, when $J_3=J_1/2$ and $J_4=J_2/2$.
  Magnified views of the separated bands are shown in (b), (c) and (d).
  Higher value of $h_z$ is assumed in (a) to show clear separation
between the bands although this phenomenon is true for any values of $h_z$.}
\label{pla_mag_R1}
\end{figure} 
\begin{figure}[h]
\includegraphics[width=260pt]{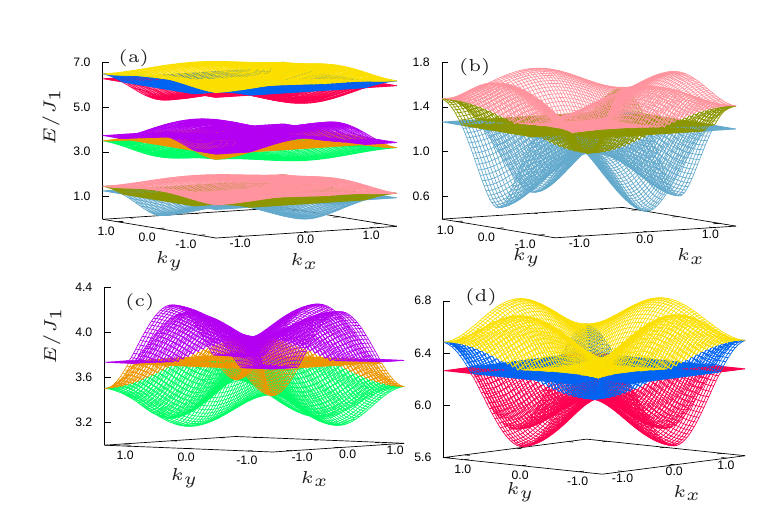}
\caption{(a) Nine triplet dispersion bands for $h_z/J_1=3.7$ and 
$J_2/J_1=1$ in R$_2$, when $J_3=J_1/2$ and $J_4=J_2/2$.
  Magnified views of the separated bands are shown in (b), (c) and (d).}
\label{pla_mag_R2}
\end{figure} 
\subsection{Effect of the magnetic field} 
The effect of magnetic field on those triplet dispersions
has been studied by applying the field along the $\hat z$ direction. 
For this purpose, the Zeeman term, $H_{\rm Z}$, has been added
to the Hamiltonian (Eq \ref{ham1}).  
$ H_{\rm Z}=h_z\sum_{i=1}^N \,S^z_{\rm T}(\boldsymbol{r}_i)$, where  
$h_z$ is the strength of the magnetic field and 
$S^z_{\rm T}(\boldsymbol{r}_i)$ is the $z$-component
of total spin of the $i$-th plaquette. 
$ H_{\rm Z}$ breaks the ${\cal T}$ and SU(2) symmetries,
but preserves the ${\cal P}$ and U(1) symmetries.
 As a result, the three-fold degeneracy of the triplet states is
 lost. 
 
In order to obtain the dispersion relations, POT has been
developed for the total Hamiltonian,
$H_{\rm T}=H \,({\rm Eq}\, \ref{ham1})+ H_{\rm Z}$. 
By expressing $H_{\rm Z}$ in terms of plaquette operators followed
by the Fourier transformation, it has the following form in the
momentum space,  
 \be
 {\cal H}_{\rm Z}=i h_z\sum_{a,\bold{k}}\left(t^\dagger_{a,\bold{k},x}t_{a,\bold{k},y} -t^\dagger_{a,\bold{k},y}t_{a,\bold{k},x}\right). 
\label{hmag}
\ee
The singlet plaquette operators, $s_{j,\bold{k}}$, do not appear
in ${\cal H}_{\rm Z}$, since the energy of singlet states
remains unaffected by the presence of magnetic field.
As a result, the singlet ground state energy does not
depend on it. 
 After the quadratic approximation, the total
 Hamiltonian becomes ${\cal H}_{\rm T}={\cal H}_{\rm Q}\,
 ({\rm Eq}\,\ref{ham3})+ {\cal H}_{\rm Z}$.
 To perform the Bogoliubov diagonalization, ${\cal H}_{\rm T}$
 has been expressed in terms of a eighteen-component vector.
 Dispersion relations are obtained numerically 
 following the Bogoliubov diagonalization of the 18$\times$18 matrix   
as described in Appendix \ref{mean}. 
 
 Every three-fold degenerate dispersion band has been splitted into
 three non-degenerate bands in such a fashion that 
 nine bands ultimately form three groups of bands in which each group 
 contains three bands. This is true for each region.
 Energy gap between the group of bands increases with the increase of $h_z$,
 but without changing the energy of the lowest group of band.
 So, the value of $\Delta$ does not change with $h_z$. 
 These dispersion bands are shown in Figs \ref{pla_mag_R1} and
 \ref{pla_mag_R2} for regions R$_1$ and R$_2$, respectively.
 Among the three, the particular
 group of non-degenerate bands having the lowest energy
 is identical with the degenerate bands in a sense that feature of 
 each of the three dispersion relation 
 is the same to that when the magnetic field was absent.
 Other two groups of non-degenerate bands have been shifted towards the
 higher energies with a little deformation in their dispersion relations.
 This deformation is perhaps due to the quadratic approximation. 
 These three groups of bands have been depicted separately 
 in Figs \ref{pla_mag_R1}-\ref{pla_mag_R2} (b), (c) and (d), 
 for the respective regions. Obviously, the mode of splitting 
 remains the same irrespective of the direction of the applied magnetic field.
 This splitting is similar to that of a triplet state under
 the magnetic field with the difference that energy values of the
 shifted states are not symmetric about that of the $S_{\rm T}^z=0$ state.
 This difference attributes to the fact that triplet states
 corresponding to $t^\dag_\alpha\ket{0}, \, \alpha=x,y$, 
 are not the eigenstates of $S_{\rm T}^z$. 
 Upon examining the structure of individual
 group of bands more closely, it reveals that all the respective 
 topological nodes and nodal-loops are there as before when the
 magnetic field was absent and their features remain unaltered.
 The features of the nodes
 and nodal-loops within each group of band
 is robust against the change of $h_z$ and
 and they cannot be destroyed by increasing the value of $h_z$.
 However, in this case, they are doubly degenerate
 and protected by both the ${\cal P}$ and U(1) symmetries.
 So, these loops cannot be termed as Dirac nodal-loops either 
 in the presence or absence of magnetic field. 
\section{Bond operator theory} 
\label{BOT}
\begin{figure}[t]
   \includegraphics [width=250pt]{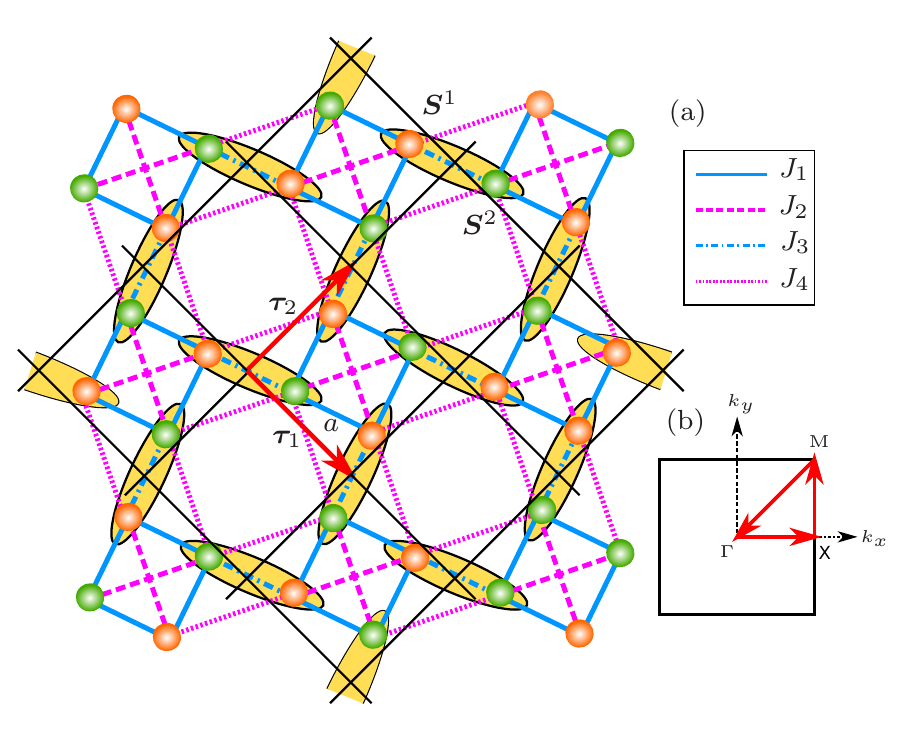}
  \caption{(a) Schematic representation of the dimerized state on
    the CaVO lattice, where the singlet dimer
    on the $J_3$ bond is shown by the narrow ellipse,
    (b) BZ of the dimerized lattices defined by $\Gamma=(0,0)$,
    X$=(\frac{2\pi}{\sqrt{10}a},0)$ and
    M$=(\frac{2\pi}{\sqrt{10}a},\frac{2\pi}{\sqrt{10}a})$.}
\label{lattice_dimer}
\end{figure}

In this section, BOT has been formulated for the
AFM Heisenberg model on the CaVO lattice, where
the system is studied in terms of weakly interacting
NN bonds connecting the plaquettes. So, in this case,
$J_3>J_1$, and the ground state is composed of
singlet dimer on the inter-plaquette NN bonds with strength $J_3$.
This ground state is known as the dimerized state, whose
geometrical view on the CaVO lattice is shown in Fig \ref{lattice_dimer} (a).
Values of $E_{\rm G}$ and $\Delta$ of the dimerized state
for this model have been determined along with the dispersion of the
triplet bond excitation. BOT has been developed by following
the formalism introduced before by Sachdev and other
in the complete basis space spanned by the singlet, $\ket{s}$,
and three triplet states, $\ket{t_{\alpha}},\, \alpha=x,y,z$,
on the bond \cite{Sachdev2}.

The bosonic singlet and triplet creation operators are defined as 
\begin{equation}
 \begin{aligned}
   & \ket{s}= s^\dagger\ket{0}=\frac{1}{\sqrt{2}}\left(\ket{\uparrow\downarrow}-\ket{\downarrow\uparrow}\right), \;\\
   & \ket{t_{x}}= t^\dagger_{x}\ket{0}= -\frac{1}{\sqrt{2}}\left(\ket{\uparrow\uparrow}-\ket{\downarrow\downarrow}\right), \;\\
   &  \ket{t_{y}}= t^\dagger_{y}\ket{0}= \frac{i}{\sqrt{2}}\left(\ket{\uparrow\uparrow}+\ket{\downarrow\downarrow}\right), \;\\
   & \ket{t_{z}}= t^\dagger_{z}\ket{0}=\frac{1}{\sqrt{2}}\left(\ket{\uparrow\downarrow}+\ket{\downarrow\uparrow}\right). \;
     \end{aligned}
 \end{equation}
The physical constraint considering the completeness relation is 
\begin{equation}
 \begin{aligned}
s^\dagger s + \sum_{\alpha} t^\dagger_{\alpha}\,t_{\alpha}=1,  
 \label{constraint_dimer}
 \end{aligned}
 \end{equation}
where the Hamiltonian for a single bond assumes the form
\[ \boldsymbol{S}^1\!\cdot\!\boldsymbol{S}^2 =-\frac{3}{4}s^\dagger s
+ \frac{1}{4} t^\dagger_{\alpha}\,t_{\alpha}. \]
The spin operators, $S^n_\alpha$, in
terms of the bond operators $s^\dagger$ and $t^\dagger_{\alpha}$ read as
 \begin{equation}
 \begin{aligned}
  S^n_\alpha=&\frac{(-1)^{n-1}}{2}\left(t^\dagger_{\alpha}\,s+ s^\dagger t_{\alpha}\right)-\frac{i}{2}\,\epsilon_{\alpha\beta\gamma}\,t^\dagger_{ \beta}\,t_{ \gamma}.
  \label{ope_dimer}
 \end{aligned}
 \end{equation}
Here, $n=1,2$, specifies the positions of two spins in a bond and 
$\alpha,\beta,\gamma=x,y,z$. The lattice generated by the middle points of every
dimer is essentially a square one, whose primitive cell may be 
constructed by the two primitive vectors, 
$\boldsymbol{\tau}_1$ and $\boldsymbol{\tau}_2$, where 
$ \boldsymbol{\tau}_1=\frac{\sqrt{10}\,a}{2}\,\hat{x}$, and 
$ \boldsymbol{\tau}_2=\frac{\sqrt{10}\,a}{2}\,\hat{y}$. 
Again, $a$ specifies the NN lattice spacing of the CaVO lattice which
 was assumed before unity. The area of the primitive cell in this case
 is one-half to the area of that used for developing the POT. 
So, area of BZ is double to that for the previous case as shown in 
 Fig \ref{lattice_dimer} (b). 
 
 The Heisenberg Hamiltonian to formulate the BOT on the CaVO lattice
 can be written as
\begin{equation}
 \begin{aligned}
 H'\!=&\sum\limits_{i}\!\big[J_3\boldsymbol{S}^1_{\boldsymbol{r}_i}\!\cdot\!\boldsymbol{S}^2_{\boldsymbol{r}_i}\!
+\!J_1\!\left(\boldsymbol{S}^2_{\boldsymbol{r}_i}\!\cdot\!\boldsymbol{S}^1_{\boldsymbol{r}_i+\boldsymbol{\tau}_1}\!+
 \boldsymbol{S}^2_{\boldsymbol{r}_i}\!\cdot\!\boldsymbol{S}^1_{\boldsymbol{r}_i+\boldsymbol{\tau}_2}\right)+\\
 &\!J_4\!\left(\boldsymbol{S}^2_{\boldsymbol{r}_i}\!\cdot\!\boldsymbol{S}^2_{\boldsymbol{r}_i+\boldsymbol{\tau}_1}\!+
 \boldsymbol{S}^1_{\boldsymbol{r}_i}\!\cdot\!\boldsymbol{S}^1_{\boldsymbol{r}_i+\boldsymbol{\tau}_2}\right)
 +\!J_2 \boldsymbol{S}^2_{\boldsymbol{r}_i}\!\cdot\!\boldsymbol{S}^2_{\boldsymbol{r}_i+\boldsymbol{\tau}_1+\boldsymbol{\tau}_2}\big].
 \end{aligned}
  \end{equation} 
Here, $\boldsymbol{S}^n_{\boldsymbol{r}_i}$ is the $n$-th spin of the $i$-th bond.
In terms of bond operators, the Hamiltonian looks like
\begin{equation}
 \begin{aligned}
 {\cal H}'&=H_1+H_{2}+H_{3},\;{\rm where}\\
 H_1&=J_3\sum\limits_{i}\left(-\frac{3}{4}s^\dagger_{\boldsymbol{r}_i}s_{\boldsymbol{r}_i}+\frac{1}{4}t^\dagger_{\boldsymbol{r}_i,\alpha}t_{\boldsymbol{r}_i,\alpha}\right)\\
 &-\mu\sum\limits_{i}\left(s^\dagger_{\boldsymbol{r}_i}s_{\boldsymbol{r}_i}+t^\dagger_{\boldsymbol{r}_i,\alpha}t_{\boldsymbol{r}_i,\alpha}-1\right),\\
 H_2=&\sum\limits_{i,m,\alpha}g(m)\Big(t^\dagger_{\boldsymbol{r}_i,\alpha}t_{\boldsymbol{r}_i+\boldsymbol{\tau'}_m,\alpha}s^\dagger_{\boldsymbol{r}_i}s_{\boldsymbol{r}_i+\boldsymbol{\tau'}_m}\\
  &+t^\dagger_{\boldsymbol{r}_i,\alpha}t^\dagger_{\boldsymbol{r}_i+\boldsymbol{\tau'}_m,\alpha}s_{\boldsymbol{r}_i}s_{\boldsymbol{r}_i+\boldsymbol{\tau'}_m}+\bold{H.c.}\Big),\\
 H_3=&\sum\limits_{i,m,\alpha}\epsilon_{\alpha\beta\gamma}\epsilon_{\alpha\beta'\gamma'}
 g'(m)t^\dagger_{\boldsymbol{r}_i,\beta}t^\dagger_{\boldsymbol{r}_i,\gamma}t_{\boldsymbol{r}_i+\boldsymbol{\tau'}_m,\beta'}t_{\boldsymbol{r}_i+\boldsymbol{\tau'}_m,\gamma'},
 \end{aligned}
  \label{ham_dimer}
  \end{equation} 
with $m=1,2,3$, $\boldsymbol{\tau}'_1=\boldsymbol{\tau}_1$,
$\boldsymbol{\tau}'_2=\boldsymbol{\tau}_2$, 
$\boldsymbol{\tau}'_3=\boldsymbol{\tau}_1+\boldsymbol{\tau}_2$,
$g(1)=g(2)=-J_1/4+J_4/4$, $g'(1)=g'(2)=J_1/4+J_4/4$, and
$g(3)=g'(3)=J_2/4$.  
Effect of the constraint, Eq \ref{constraint_dimer},
has been taken into account in Eq \ref{ham_dimer} like before. 
\subsection{Quadratic and mean-field approximations}
In quadratic approximation, contribution of $H_{3}$ is neglected.
So, the Hamiltonian in the momentum space
is, ${\cal H}'_{\rm Q}=E'_0+H_{2}$,  
where 
\begin{equation}
  \begin{aligned}
  E'_0=&N^\prime\left[-\frac{3}{4}J_3\bar s^2-\mu\left(\bar s^2-1\right)\right],\\
 H_{2}=&\sum\limits_{\bold{k}} A_{\bold{k}} t^\dagger_{\bold{k},\alpha}t_{\bold{k},\alpha}
  + \frac{B_{\bold{k}}}{2}\!\left(t^\dagger_{\bold{k},\alpha}t^\dagger_{-\bold{k},\alpha}+
  t_{-\bold{k},\alpha}t_{\bold{k},\alpha}\right),
  \end{aligned}
  \label{H2_dimer}
 \end{equation}
   with $N^\prime=2N$, the total number of dimers in the system and 
 \begin{equation}
  \begin{aligned}
 A_{\bold{k}}&=(\frac{J_3}{4}-\mu)+B_{\bold{k}},\\
 B_{\bold{k}}&=2\,\bar s^2 \sum_{m=1,2,3}g(m)\,\cos\left(\bold{k}\cdot \boldsymbol{\tau}'_m\right). \nonumber
 \end{aligned}
 \end{equation}
Condensation of the singlets is implemented by the substitution, 
$s^\dagger_{\boldsymbol{r}_i}=s_{\boldsymbol{r}_i}=\langle s^\dagger_{\boldsymbol{r}_i}\rangle
=\langle s_{\boldsymbol{r}_i}\rangle=\bar s$ \cite{Sachdev2}.
The values of $\mu$ and $\bar s^2$ are determined by
solving the pair of following self-consistent equations. 
\begin{equation}
  \begin{aligned}
&\mu=-\frac{3}{4}J_3+\frac{3}{2N'}\sum\limits_{\bold{k}}
\left[\frac{( A_{\bold{k}}- B_{\bold{k}})} {\Omega_{\bold{k}}}-1\right]\frac{ B_{\bold{k}}}{\bar s^2}, \\
& \bar s^2=1+\frac{3}{2N'}\sum\limits_{\bold{k}}
\left[1-\frac{ A_{\bold{k}}}{\Omega_{\bold{k}}}\right].
\label{selfcon1}
  \end{aligned}
\end{equation}
Diagonalizing the Hamiltonian, $H_2$ in Eq \ref{H2_dimer}, by the
bosonic Bogoliubov transformation, triplet dispersion is obtained,
which is \(\Omega_{\bold{k}}=\sqrt{A^2_{\bold{k}}-B^2_{\bold{k}}}\), along with 
the ground state energy of the system in BOT, 
\(E'_{\rm G}=E'_0+\frac{3}{2}\sum_{\bold{k}}
\left(\Omega_{\bold{k}}-A_{\bold{k}}\right)\).
  
To obtain more accurate value of $E_{\rm G}$,
contribution of the terms containing quartic triplet operators,
$H_3$ in Eq \ref{ham_dimer}, is taken into account
by performing mean-field approximation on them. 
The terms of cubic order in the triplet operators do
not contribute since the condensation of triplet operators
is not allowed in this formulation \cite{Sachdev2}. 
By introducing the real space mean-field order parameters, 
$P(m)=\sum_{\beta}\langle t^\dagger_{{\boldsymbol{r}_i},\beta}
t_{{\boldsymbol{r}_i+\boldsymbol{\tau'}_m},\beta}\rangle$ and 
$Q(m)=\sum_{\beta}\langle t^\dagger_{{\boldsymbol{r}_i},\beta} t^\dagger_{{\boldsymbol{r}_i+\boldsymbol{\tau'}_m},\beta}\rangle$ with $m=1,2,3$,  
the mean-field Hamiltonian \cite{Susobhan}, $H_{\rm MF}$ becomes, 
\begin{equation}
 \begin{aligned}
 H_{\rm MF}&=H'_1+H'_{2},\, {\rm with}\;\\
 H'_1&=H_1+\frac{2}{3}\sum\limits_{m}g'(m)\left[Q^2(m)-P^2(m)\right],\\
 H'_2&=H_2+\frac{2}{3}\,\sum\limits_{i,m}g'(m)P(m)\left(t^\dagger_{\boldsymbol{r}_i,\alpha}t_{\boldsymbol{r}_i+\boldsymbol{\tau'}_m,\alpha}+\bold{H.c.}\right)\\
 &-\frac{2}{3}\,\sum\limits_{i,m}g'(m)Q(m)\left(t^\dagger_{\boldsymbol{r}_i,\alpha}t^\dagger_{\boldsymbol{r}_i+\boldsymbol{\tau'}_m,\alpha}+\bold{H.c.}\right).
\end{aligned}
  \end{equation}  
Geometrical symmetries of $\boldsymbol{\tau}'_m$s allow to consider
the following relations, $P(1)=P(2)=P_1$, $P(3)=P_2$, $Q(1)=Q(2)=Q_1$
and $Q(3)=Q_2$.
In the momentum space, $H_{\rm MF}=E'_0+H'_{2}$, where, 
\begin{equation}
  \begin{aligned}
    E^\prime_0=&N^\prime\left[-\frac{3}{4}J_3\bar s^2-\mu\left(\bar s^2-1\right)\!
      +\!\frac{1}{2}\sum\limits_{j=1,2}r_j\left(Q^2_j-P^2_j\right)\right],\\
    H'_{2}=&\sum\limits_{\bold{k}} A^\prime_{\bold{k}} t^\dagger_{\bold{k},\alpha}
    t_{\bold{k},\alpha}
 +\frac{B^\prime_{\bold{k}}}{2}\!\left(t^\dagger_{\bold{k},\alpha}t^\dagger_{-\bold{k},\alpha}
 +  t_{-\bold{k},\alpha}t_{\bold{k},\alpha}\!\right),
     \end{aligned}
 \end{equation}
 with 
 $r_1=4\left(g^\prime(1)+g^\prime(2)\right)/3$, $r_2=4g^\prime(3)/3$ and 
$A^\prime_{\bold{k}}=A_{\bold{k}}+\frac{4}{3}\sum_{m}g'(m)P(m)\cos\left(\bold{k}\cdot \boldsymbol{\tau}'_m\right)$,  
 $B^\prime_{\bold{k}}=B_{\bold{k}}-\frac{4}{3}\sum_{m}g'(m)Q(m)\cos\left(\bold{k}\cdot \boldsymbol{\tau}'_m\right)$.
 Values of the order parameters have been determined
 by solving two pairs of self-consistent equations,   
\begin{equation}
  \begin{aligned}
&P_j=\frac{3}{2}\,\frac{1}{N'r_j}\sum\limits_{\bold{k}}
\left[\left(\frac{A'_{\bold{k}}}{\Omega_{\bold{k}}}-1\right)\frac{\partial A'_{\bold{k}}}{\partial P_j}\right], \\
&Q_j=\frac{3}{2}\,\frac{1}{N'r_j}\sum\limits_{\bold{k}}
\left[\frac{B'_{\bold{k}}}{\Omega_{\bold{k}}}\frac{\partial B'_{\bold{k}}}{\partial Q_j}\right],
\label{selfcon2}
  \end{aligned}
\end{equation}
with $j=1,2$, in addition to the equations those are previously
obtained for $\mu$ and $\bar s^2$, in Eq \ref{selfcon1}.  
Diagonalizing the Hamiltonian $H'_2$ like before,  
the mean-filed dispersion relation,
$\Omega'_{\bold{k}}=\sqrt{A^{\prime2}_{\bold{k}}-B^{\prime2}_{\bold{k}}}$, 
and the mean-field ground state energy, $E''_{\rm G}=E'_0
+\frac{3}{2}\sum_{\bold{k}}
\left(\Omega^\prime_{\bold{k}}-A^\prime_{\bold{k}}\right)$, 
have been obtained.  
Ground state energy per bond has been obtained by 
using the second-order perturbation theory, which has the expression,      
$E^{\rm P}_{\rm G}=-3J_3\left(1+\left((J_1-(J_2/2)-J_4)/2
\sqrt 2 J_3\right)^2\right)/4$.
However, in this case, the unperturbed Hamiltonian is
the sum of all the NN inter-plaquette interactions, 
$H'_0=J_3\sum_{i=1}^{N'}\boldsymbol{S}^1_{\boldsymbol{r}_i}
\cdot\boldsymbol{S}^2_{\boldsymbol{r}_i}$,
while the perturbation, $H'-H'_0$, includes all the remaining terms.
This result is valid 
as long as the strengths of NN inter-plaquette interactions are
stronger than others.
\begin{figure}[h]
 \begin{center}
   \includegraphics[width=250pt]{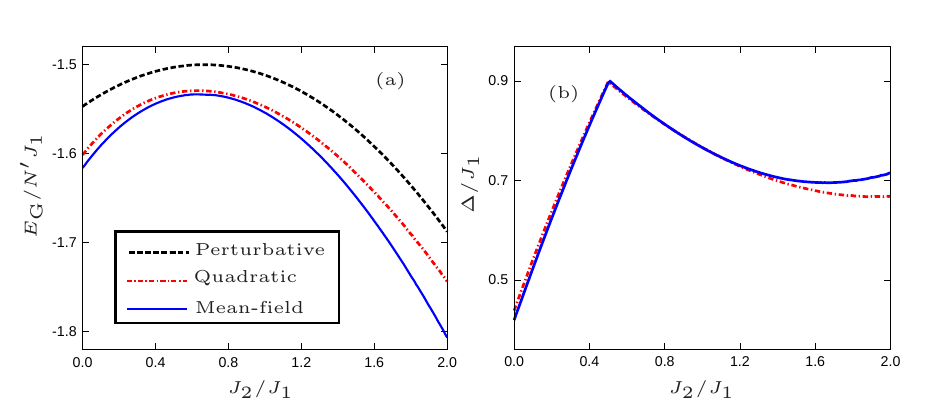}
 \end{center}
 \caption{Variation of (a) Ground-state energy per bond and
   (b) spin gap against $J_2/J_1$ for $J_3=2J_1$ and $J_4=J_2$.}
     \label{Edimer}
  \end{figure}   
Variation of ground state energy per bond for the
regime, $0<J_2/J_1<2$, obtained in quartic,
mean-field and perturbative approximations with respect to
$J_2/J_1$ are shown in Fig \ref{Edimer} (a), for $J_3=2J_1$ and $J_4=J_2$.
Mean-field estimation, $E''_{\rm G}$ is the lowest since it includes the
contributions of the quartic terms. 
\begin{figure}[h]
\begin{center}
  \includegraphics[width=150pt]{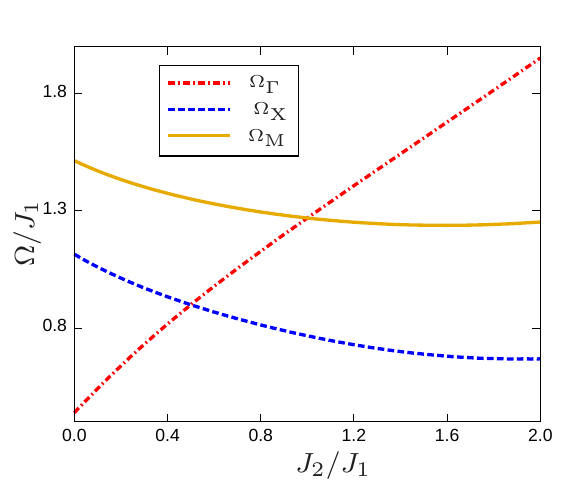}
 \end{center}
   \caption{ Variation of energies of  
 $ \Omega_{\Gamma}$, $ \Omega_{{\rm M}}$ and $ \Omega_{{\rm X}}$ 
     measured with respect to $E'_{\rm G}$ with $J_2/J_1$,
     when $J_3=2J_1$ and $J_4=J_2$.}
\label{gapdimer}
\end{figure}

In order to estimate the value of $\Delta$, variations of the
triplet energies for the high-symmetry points,
$\Omega_{\Gamma}$, $\Omega_{{\rm X}}$ and $\Omega_{{\rm M}}$
have been plotted with $J_2/J_1$, when $J_3=2J_1$ and $J_4=J_2$ in 
Fig \ref{gapdimer}, as the minima of $\Omega_{\bold{k}}$
are found to occur at those points. 
It shows that $\Omega_{\Gamma}$ and  $\Omega_{{\rm X}}$ cross 
each other at the point $J_2/J_1=1/2$, in such a way that 
 $\Omega_{\Gamma}$ is the lowest when $J_2/J_1<1/2$
while $\Omega_{{\rm X}}$ is that when $J_2/J_1>1/2$.
The value of triplet gap, $\Delta$, which accounts the separation between 
ground and the lowest triplet state energies has been obtained
for the regime, $0<J_2/J_1<2$. Variation of $\Delta$
is shown in Fig \ref{Edimer} (b). 
\begin{figure}[h]
       \centering
         \includegraphics[width=219pt]{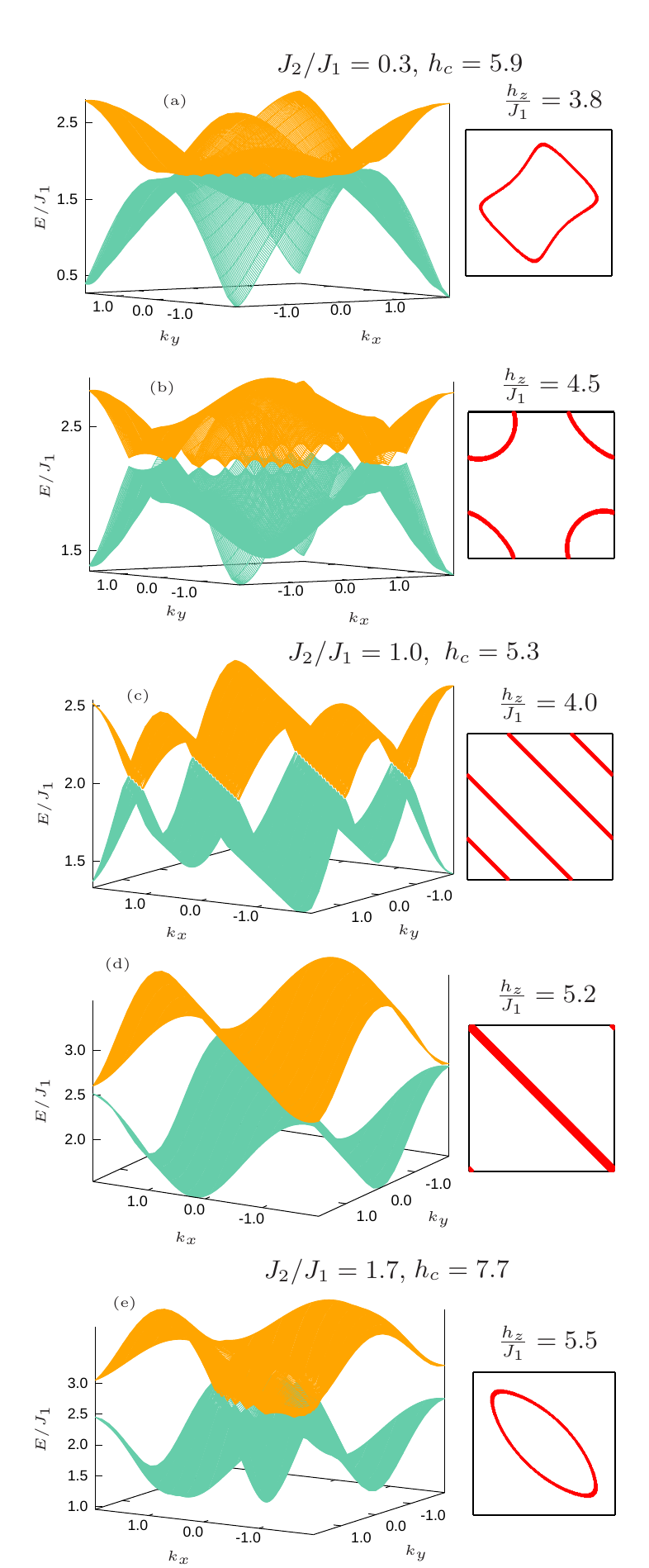}
  \caption{Forms of nodal-loops and nodal-lines
    for different values of interaction strengths and
    magnetic fields when $J_3=2J_1$ and $J_4=J_2$.
    The topmost triplon band is not shown as it is
  clearly separated from the middle band in every case.}
     \label{mag}
\end{figure}
To investigate the effect of magnetic field,
${\cal H}'_{\rm Z}$ has been added to ${\cal H}'$. 
Expression of ${\cal H}'_{\rm Z}$ in terms of triplet operators
in the momentum space is, 
${\cal H}'_{\rm Z}=i h_z\sum_{\bold{k}}\left(t^\dagger_{\bold{k},x}t_{\bold{k},y}
-t^\dagger_{\bold{k},y}t_{\bold{k},x}\right)$, when the magnetic field acts along
the $\hat z$ direction.
Now, the degenerate triplet band splits into three
non-degenerate triplon bands and the separation between them increases with
the increase of $h_z$. They are completely separated from each other
above the critical values of magnetic field, $h_c$,
where the value of $h_c$ depends on
the values of the exchange parameters.
Magnetic field induced nodal-lines
of various forms and positions on the BZ are found within the
lower triplon dispersion bands, as long as $h_z<h_c$. For examples, 
nodal-loops are found when $J_2/J_1=0.3$ and 1.7,
for $J_3=2J_1$ and $J_4=J_2$, which
are shown in Figs \ref{mag} (a), (b) and (e).
Elliptic loop is noticed when $J_2/J_1=1.7$ and $h_z/J_1<7.7$. 
Closed loop with various shapes can be obtained for $J_2/J_1=0.3$
by changing the values of $h_z$ as long as $h_z/J_1<5.9$. 
Straight nodal-lines are obtained when $J_2/J_1=1$, for $4.0<h_z/J_1<5.2$
(Figs \ref{mag} (c) and (d)). 

So, all of those doubly-degenerate nodal-lines are protected by 
the ${\cal P}$ and U(1) symmetries. 
These magnetic field induced nodal-lines do not survive
above the $h_c$. It is worthy to state that
the dimerized ground state is unstable at the higher magnetic field.
So the validity of the BOT is questionable in the presence of
high magnetic field. 
\section{Discussion}
\label{Discussion}
In this study, emergence of topological nodal-lines with various features 
has been reported in the frustrated AFM spin-1/2 Heisenberg model
formulated on the CaVO lattice. CaVO lattice can be transformed into
two different square lattices by treating either 
four-site plaquettes or two-site bonds as a basis sites
as shown in Figs  \ref{lattice} (c)
and \ref{lattice_dimer} (a), respectively. POT and BOT have been developed in
two different parameter regimes by choosing the values of
NN and NNN inter- and intra-plaquette interactions 
in such a way that the effective Hamiltonians in terms of
plaquette and bond operators
formulated on the respective square lattices are valid. 
Dispersion relations of triplet plaquette and bond excitations,
based on the PRVB and dimerized ground states are obtained, where
nodal-lines are found to exist in the presence and absence of magnetic field. 
Ground state energy and spin gap in both the regimes have been
obtained in this context. 

Emergence of a pair of six-fold degenerate nodal-loops with circular
and square shapes are noted
in the plaquette dispersions, which are ${\cal PT}$ and SU(2) protected.
In the presence of magnetic field, three pairs of
two-fold degenerate nodal-loops 
of almost the same features are found, those are ${\cal P}$ and U(1) protected.
The system hosts a number of magnetic field induced nodal-lines
of various shapes in the triplon dispersions.
The system hosts only non-Dirac nodal-lines
because of the fact that none of them are four-fold degenerate
and protected by the ${\cal PT}$ symmetry. 
But a pair of Dirac nodal-loops have been
noted before in the AFM magnon dispersions for a 
Heisenberg model on this lattice
by considering a larger unit cell containing eight sites \cite{Owerre}.
This difference attributes to the fact that AFM ground state breaks the
full symmetry of the Heisenberg Hamiltonian, while the singlet
ground states in this study do not.
Emergence of multiple topological phases in irradiated tight-binding and
FM Kitaev-Heisenberg models 
on this lattice is noted before \cite{Arghya,Moumita2}. 
The effect of DMI cannot be registered either in
POT or BOT because of the fact that, $\sum_n ({\boldsymbol S}^n \times
{\boldsymbol S}^{n+1})_z$ vanishes identically over
a plaquette and bond, when
$\boldsymbol S$ is expressed in terms of either plaquette or
bond operators, respectively.

The value of spin-gap for CaV$_4$O$_9$ has been determined by
measuring the uniform spin susceptibility, and the NMR spin-lattice
relaxation rate \cite{Taniguchi}, as well as by the
inelastic neutron scattering \cite{Kodama}. In the scattering experiment,
both powder sample and single crystal of CaV$_4$O$_9$ have been used. 
Both $Q$- and $\theta$-scans have been performed where $Q$ and $\theta$
are the absolute value of scattering wave vector and sample-angle,
respectively. For the measurement of spin-gap only the lowest energy
dispersion branch has been determined.

However, in order to
observe the nodal points and lines, at least a pair of
lower energy dispersion branches are to be determined distinctly 
in the inelastic neutron scattering experiment. For this purpose,
wave vector dependent scans on the single crystal would be useful.
Repeated scans with varying scattering wave vector
in the presence of magnetic field are necessary, however, 
the direction of the field is irrelevant.
BOT predicted several types of nodal-lines (Fig \ref{mag}) only
in the presence of magnetic field, while, POT noted the circular
nodal-loop (Fig \ref{nodal_ring}) between two lower bands
where those lines occur at a fixed energy.
So, scan with fixed incident neutron energy is sufficient for
there observation. On the other hand, 
the square nodal-loop (Fig \ref{nodal_line})
found in POT occurs between the upper two bands
where it is spanned over a finite energy range. 
Scanning with incident neutrons with a wide range of
energies is thus required for its detection.  

 \section{ACKNOWLEDGMENTS}
MD acknowledges the UGC fellowship, No. 524067 (2014), India.  
 \section{Author contribution statement}
 MD did both analytical and numerical works as well as
 prepared the figures. 
Both the authors were involved in the preparation of the
manuscript. 
 \appendix
 \section{Details of plaquette operator theory}
  \label{mean}
\subsection{Symmetries of eigenstates of four-spin Heisenberg plaquette}
\label{eigensystem}
Symmetries of all the eigenvectors of the single square
plaquette, $H_{\text{\Square}}$, are described
in terms of eight operators of dihedral group $D_4$,
four rotations, $R_n$ and reflections, $M_n$, $n=1,2,3,4$.
Here, $R$ implies the rotation by $\pi/2$ about the center of the square,
as depicted in Fig  \ref{lattice} (e) and $R_n$ means
successive $R$ operation by $n$ times.
Rotation $R$ can be defined as  
$R\ket {S_1S_2S_3S_4}=\ket{S_2S_3S_4S_1}$, where 
$\ket{S_1S_2S_3S_4}=\ket{S_1^z}\otimes\ket{S_2^z}\otimes\ket{S_3^z}
\otimes\ket{S_4^z}$, in which 
$\ket{S_n^z}$ is the spin state at the $n$-th vertex of the square.
Rotational property of an eigenstate, $|\nu\rangle$,    
can be studied in terms of an eigenvalue equation, 
$R_p|\nu\rangle=\lambda_r|\nu\rangle$, where  
$\lambda_r=\pm 1$ be the eigenvalue of the rotational operator $R_p$.   
Where $p<4$ and it corresponds that minimum number of $R$
operations on $|\nu\rangle$ for which 
$\lambda_r$ assumes the value either $+1$ or $-1$. 
$\lambda_r=\pm 1$ imply the even (symmetric) 
and odd (antisymmetric) parity of the state. 
Each state has definite parity and
among all six are symmetric. 
It is found that 
$p$ assumes the value  either 1 or 2.
For $\Psi_{\rm RVB}$ ($\Psi_{\rm aRVB}$), $\lambda_r=1(-1)$ and $p=1$,
which means that
$\Psi_{\rm RVB}$ is symmetric, whereas, 
$\Psi_{\rm aRVB}$ is antisymmetric under the rotation by the angle $\pi/2$. 

Similarly, reflectional symmetry of $|\nu\rangle$ has been studied in 
terms of an eigenvalue equation 
$M_n|\nu\rangle=\lambda_{M_n}|\nu\rangle$, 
where four different mirror planes for 
$M_n$ are shown by dashed lines in Fig \ref{lattice} (f).   
${ M}_n$ are defined as 
${M}_1\ket {S_1S_2S_3S_4} =\ket{S_1S_4S_3S_2},\;
{M}_2\ket {S_1S_2S_3S_4} =\ket{S_3S_2S_1S_4},\;
{M}_3\ket {S_1S_2S_3S_4}=\ket{S_2S_1S_4S_3},\;
{M}_4\ket {S_1S_2S_3S_4}=\ket{S_4S_3S_2S_1}$.
Values of $\lambda_{ M_3}$ and $\lambda_{ M_4}$
remain undefined for the degenerate triplets, $\ket{t_{1}}$
and $\ket{t_{2}}$. However, values of $p$, $\lambda_r$ and $\lambda_{M_n}$
for all the states are given in Table I. 
Again, $\Psi_{\rm RVB}$ is symmetric, while  
$\Psi_{\rm aRVB}$ is antisymmetric under the reflection about the 
mirror planes either $M_1$ or $M_2$ as shown in Fig \ref{lattice} (f). 
But, both $\Psi_{\rm RVB}$ and $\Psi_{\rm aRVB}$ are found symmetric
under spin inversion as well as 
reflection about the mirror planes,  
$M_3$ and $M_4$. 
\begin{table}[h]
\begin{center}
\caption{Energy and other eigenvalues of the eigenstates of spin-1/2  Heisenberg square plaquette}
\def\arraystretch{0.78}
  \tabcolsep4pt\begin{tabular}{|c|c|c|c|c|c|c|c|}
\hline
 $\boldsymbol{S_{\rm T}}$  &\textbf{Energy eigenvalues}& $\boldsymbol{\lambda_r}$ & $\boldsymbol{p}$ & $\boldsymbol{\lambda_{ M_1}}$ & $\boldsymbol{\lambda_{ M_2}}$ & $\boldsymbol{\lambda_{ M_3}}$ & $\boldsymbol{\lambda_{ M_4}}$ \\  
\hline
 0  & $ E_{s_{1}}=  -2J_1+\frac{1}{2}J_2 $ & 1 & 1 &1 &1 &1 &1 \\
\hline
0  & $ E_{s_2}=-\frac{3}{2}J_2 $  &- 1 & 1 &-1 &-1 &1 &1 \\
\hline
1   &$E_{t_1}=-\frac{1}{2}J_2$ & -1 &2&1 &-1 & & \\
\hline
1  &$E_{t_{ 2}}  =  -\frac{1}{2}J_2$& -1 & 2&-1 &1 & & \\
\hline
1  & $E_{t_{3}}  =  -J_1+\frac{1}{2}J_2$& -1 & 1 &1 & 1&-1 &-1 \\
\hline
2  & $ E_{q}= J_1+\frac{1}{2}J_2$ & 1 & 1&1 &1 &1 &1\\
\hline
\end{tabular}
\end{center}
\label{table_eigenvalues}
\end{table}
In order to express the eigenstates in a compact form 
following notations are used.
\begin{equation}
\begin{aligned}
 & \ket{\psi^2_n}=T^{n-1}\ket{2}\left(n=1\right), \ket{2}=\ket{\uparrow\uparrow\uparrow\uparrow},\\
 &\ket{\psi^1_n}=T^{n-1}\ket{1}\left(n=1,2,3,4\right), \ket{1}=\ket{\downarrow\uparrow\uparrow\uparrow}, \\
  & \ket{\psi^0_n}_1=T^{n-1}\ket{0}_0\left(n=1,2,3,4\right), \ket{0}_0=\ket{\uparrow\uparrow\downarrow\downarrow}, \\
   & \ket{\psi^0_n}_2=T^{n-1}\ket{0}_3\left(n=1,2\right), \ket{0}_3=\ket{\uparrow\downarrow\uparrow\downarrow}, \\
 & \ket{\psi^{-1}_n}=T^{n-1}\ket{-1}_1\left(n=1,2,3,4\right), \ket{-1}_1=\ket{\uparrow\downarrow\downarrow\downarrow}, \\
  & \ket{\psi^{-2}_n}=T^{n-1}\ket{-2}\left(n=1\right), \ket{-2}=\ket{\downarrow\downarrow\downarrow\downarrow}.\nonumber
   \label{bra}
 \end{aligned}
\end{equation}
 Here $T$ operator is defined as, 
 $T\ket{pqrs}=\ket{spqr}$, where $\ket{pqrs}=\ket{p}\otimes\ket{q}\otimes\ket{r}\otimes\ket{s}$.
All the energy eigenstates are written below.
 \begin{equation}
 \begin{aligned}
\ket{s_1} &=\frac{1}{\sqrt{12}}\left(2\sum\limits_{n=1}^{2} \ket{\psi^0_n}_2- \sum\limits_{n=1}^{4}\ket{\psi^0_n}_1\right), \\
\ket{s_2}&=\frac{1}{2}\sum\limits_{n=1}^{4}\left(-1\right)^{n} \ket{\psi^0_n}_1, \nonumber\\
\ket{t_{1,\alpha}}&=\frac{\lambda_{\alpha}\lambda_{n}}{2}\sum\limits_{n=1,3}\left(\ket{\psi^1_n}\pm \ket{\psi^{-1}_n}\right),\\
\ket{t_{1,z}}&=\frac{1}{2}\left(\sum\limits_{n=2,3}\ket{\psi^0_n}_1-\sum\limits_{n=1,4}\ket{\psi^0_n}_1\right),\\
\ket{t_{2,\alpha}}&=\frac{\lambda_{\alpha}\lambda_{n}}{2}\sum\limits_{n=2,4}\left(\ket{\psi^1_n}\pm \ket{\psi^{-1}_n}\right),\\
\ket{t_{2,z}}&=\frac{1}{2}\left(\sum\limits_{n=3}^{4}\ket{\psi^0_n}_1-\sum\limits_{n=1}^{2}\ket{\psi^0_n}_1\right),\\
\ket{t_{3,\alpha}}&= \frac{\lambda_{\alpha}}{2\sqrt{2}}\sum\limits_{n=1}^{4}\left(-1\right)^{n-1}\left(\ket{\psi^1_n}\pm \ket{\psi^{-1}_n}\right),\\
\ket{t_{3,z}}&=\frac{1}{\sqrt{2}}\sum\limits_{n=1}^{2}\left(-1\right)^{n-1}\ket{\psi^0_n}_2,\\
\ket{q_{\alpha}}&=\frac{\lambda_{\alpha}}{2\sqrt{2}}\sum\limits_{n=1}^{4}\left(\ket{\psi^1_n}\pm \ket{\psi^{-1}_n}\right),\\
 \nonumber
  \end{aligned}
 \end{equation}
 \begin{equation}
 \begin{aligned}
\ket{q_z} &=\frac{1}{\sqrt{6}}\left(\sum\limits_{n=1}^{4} \ket{\psi^0_n}_1+ \sum\limits_{n=1}^{2}\ket{\psi^0_n}_2\right), \\
\ket{q_{\pm}} &=\frac{1}{\sqrt{2}}\left(\ket{\psi^2_n}\pm \ket{\psi^{-2}_n}\right), \nonumber
  \end{aligned}
     \end{equation}     
  where the upper and lower signs respectively refer to $\alpha = x$
and $y$, $\lambda_x =1$ and $\lambda_y = i$.
$\lambda_n=-1$ for $n=1,2$ and $\lambda_n=1$ for $n=3,4$.

\subsection{Terms used in four-spin plaquette operators}
The analytic expressions of the coefficients $A^n_{a}$, $B^n_{a}$
and $D^n_{ab}$ are given here,
  \begin{equation}
 \begin{aligned}
 & A^{1/3}_{1}=\pm \frac{1}{\sqrt{12}},\;
  A^{2/4}_{2}=\pm\frac{1}{\sqrt{12}},\;
  A^{1/3}_{3}=-A^{2/4}_{3}=\frac{1}{\sqrt{6}}, \\
 & B^{2/4}_{1}=\pm \frac{1}{2},\;
  B^{1/3}_{2}=\pm \frac{1}{2},\;
   D^{2/4}_{11}=D^{1/3}_{22}= \frac{1}{2}, \\
  & D^{2/4}_{33}=D^{1/3}_{33}= \frac{1}{4},\;
   D^{1/3}_{13}=-D^{2/4}_{23}= \pm\frac{1}{2\sqrt{2}}.\nonumber
  \label{lambda}
 \end{aligned}
 \end{equation}
\normalsize
Explicit forms of the terms, $H_{nm}$ of the Eq \ref{ham2}, 
in the momentum space are given here. 
  \begin{equation}
    \begin{aligned}
    H_{02}&=\sum\limits_{\bold{k}}\left(E_{s_j}-
    \mu\right)s^\dagger_{j,\bold{k}}s_{j,\bold{k}},\\
    H_{20}\!&=   \sum\limits_{\bold{k}} X^{ab}_{\bold{k}} \;t^\dagger_{a,\bold{k},\alpha}t_{b,\bold{k},\alpha} \\
    &+ \frac{1}{2}\sum\limits_{\bold{k}}Y^{ab}_{\bold{k}}\!\left(t^\dagger_{a,\bold{k},\alpha}t^\dagger_{b,-\bold{k},\alpha}+
  t_{a,-\bold{k},\alpha}t_{b,\bold{k},\alpha}\!\right), \\
  H_{30}\!&=\frac{\epsilon^{\alpha\beta\gamma}}{\sqrt{N}}\!\!\!\sum\limits_{\bold{p},\bold{k},a,b,c}\!\!\!\! \!Z^{abc}_{\bold{p}-\bold{k}} \,t^\dagger_{a,\bold{k}-\bold{p},\alpha}t^\dagger_{b,\bold{p},\beta}t_{c,\bold{k},\gamma}+\bold{H.c.}, \\
  H_{40}\!&=\frac{\epsilon^{\alpha\beta\gamma}\epsilon^{\alpha\lambda\nu}}{N}
  \!\!\!\!\!\!\!\!\!\sum\limits_{\bold{p},\bold{q},\bold{k},a,b,c,d}\!\!\!\!\!\!\!\!\!M^{abcd}_{\bold{k}}\,t^\dagger_{a,{\bold{p}}+{\bold{k}},\beta}t^\dagger_{b,{\bold{q}}-{\bold{k}},\lambda}t_{c,{\bold{q}},\nu}t_{d,{\bold{p}},\gamma},\\
   H_{21}\!&=\frac{1}{\sqrt{N}}\sum\limits_{\bold{p},\bold{k},a,b}\Big[W^{ab}_{\bold{p}}t^\dagger_{a,{\bold{k}}-{\bold{p}},\alpha}t^\dagger_{b,\bold{p},\beta}s_{j,\bold{k}}+\bold{H.c.} \\
   & + W^{ba}_{\bold{p}}s^\dagger_{j,{\bold{k}}-{\bold{p}}}t^\dagger_{a,\bold{p},\alpha}t^\dagger_{b,\bold{k},\alpha}+\bold{H.c.}\Big],\\
   H_{22}&=\frac{1}{N}\sum\limits_{\bold{p},\bold{q},\bold{k},a,b}\Big[N^{ab}_{\bold{k}}s^\dagger_{j,{\bold{q}}+{\bold{k}}}s_{j,{\bold{p}}+{\bold{k}}}t^\dagger_{a,\bold{p},\alpha}t_{b,\bold{q},\alpha}\\
  &+\frac{1}{2}N^{ab}_{\bold{k}}s^\dagger_{j,{\bold{q}}+{\bold{k}}}s^\dagger_{j,{\bold{p}}-{\bold{k}}}t_{a,\bold{p},\alpha}t_{b,\bold{q},\alpha}+\bold{H.c.}\Big],\nonumber
     \end{aligned}
 \end{equation}
 The analytic expressions of several terms are given as
     \begin{equation}
  \begin{aligned}
   X^{ab}_{\bold{k}}&=\left(E_{t_a}-\mu\right)\left(\delta_{a,1}\delta_{b,1}+\delta_{a,2}\delta_{b,2}+\delta_{a,3}\delta_{b,3}\right)
   +  Y^{ab}_{\bold{k}},\\
    Y^{ab}_{\bold{k}}&=\bar s^2\sum\limits_{m=1,2}2g^{ab}(m)\cos\left(\bold{k}\cdot \boldsymbol{\tau}_{m}\right), \\
  g^{ab}(m)&=J_1\left(A^2_aA^4_b\delta_{m,1}+A^1_aA^3_b\delta_{m,2}\right)\\
  &+J_2\bigg((A^2_aA^1_b+A^3_aA^4_b)\delta_{m,1} \\
  &+\left(A^2_aA^3_b+A^1_aA^4_b\right)\delta_{m,2}\bigg). \nonumber
   \end{aligned}
  \end{equation}
     Here $a,b=1,2,3$, and $E_{t_a}$ is the triplet energy
     of the single plaquette.  
 The expressions of all $g$ coefficients are given for the regions R$_1$ only. The expressions will be the same for region R$_2$  with the replacements $A\leftrightarrow B$.
   \begin{equation}
  \begin{aligned}
  Z^{abc}_{\bold{k}}&=i\bar s \sum\limits_{m=1,2}\left[g^{abc}_Z(m)e^{-i\bold{k}\cdot \boldsymbol{\tau}_m}+g^{bca}_{\bar Z}(m)e^{i\bold{k}\cdot \boldsymbol{\tau}_m}\right],\\
  M^{abcd}_{\bold{k}}&=-\frac{1}{2}  \sum\limits_{m=1,2}\left[g^{abcd}_M(m)e^{-i\bold{k}\cdot \boldsymbol{\tau}_m}+g^{cdab}_M(m)e^{i\bold{k}\cdot \boldsymbol{\tau}_m}\right],\\
  W^{ab}_{\bold{k}}&=\bar s  \sum\limits_{m=1,2}\left[g^{ab}_W(m)e^{-i\bold{k}\cdot \boldsymbol{\tau}_m}+g^{ba}_{\bar W}(m)e^{i\bold{k}\cdot \boldsymbol{\tau}_m}\right],\\
   N^{ab}_{\bold{k}}&= \sum\limits_{m=1,2}\left[g^{ab}_N(m)e^{-i\bold{k}\cdot \boldsymbol{\tau}_m}+g^{ba}_N(m)e^{i\bold{k}\cdot \boldsymbol{\tau}_m}\right].\nonumber
   \end{aligned}
   \end{equation}
  The $g$ coefficients are given by
  \begin{equation}
  \begin{aligned}
  g^{abc}_{ Z}&(m)=J_1\left(A^2_aD^4_{bc}\delta_{m,1}+A^1_aD^3_{bc}\delta_{m,2}\right)\\
 +&J_2\bigg((A^2_aD^1_{bc}+A^3_aD^4_{bc})\delta_{m,1} 
  +\left(A^2_aD^3_{bc}+A^1_aD^4_{bc}\right)\delta_{m,2}\bigg),\\
  g^{abc}_{\bar Z}&(m)=J_1\left(D^2_{ab}A^4_c\delta_{m,1}+D^1_{ab}A^3_c\delta_{m,2}\right)\\
 +&J_2\bigg((D^2_{ab}A^1_c+D^3_{ab}A^4_c)\delta_{m,1} 
  +\left(D^2_{ab}A^3_c+D^1_{ab}A^4_c\right)\delta_{m,2}\bigg),\\  
   g^{abcd}_M&(m)=J_1\left(D^2_{ab}D^4_{cd}\delta_{m,1}+D^1_{ab}D^3_{cd}\delta_{m,2}\right)\\
  +J_2&\bigg((D^2_{ab}D^1_{cd}\!+\!D^3_{ab}D^4_{cd})\delta_{m,1} 
 \! +\!\left(D^2_{ab}D^3_{cd}\!+\!D^1_{ab}D^4_{cd}\right)\delta_{m,2}\bigg),\\
  g^{ab}_W&(m)=J_1\left(B^2_aA^4_b\delta_{m,1}+B^1_aA^3_b\delta_{m,2}\right)\\
 & +J_2\bigg((B^2_aA^1_b+B^3_aA^4_b)\delta_{m,1} 
  +\left(B^2_aA^3_b+B^1_aA^4_b\right)\delta_{m,2}\bigg),\\
   g^{ab}_{\bar W}&(m)=J_1\left(A^2_aB^4_b\delta_{m,1}+A^1_aB^3_b\delta_{m,2}\right)\\
 & +J_2\bigg((A^2_aB^1_b+A^3_aB^4_b)\delta_{m,1} 
  +\left(A^2_aB^3_b+A^1_aB^4_b\right)\delta_{m,2}\bigg),\\
  g^{ab}_N&(m)=J_1\left(B^2_aB^4_b\delta_{m,1}+B^1_aB^3_b\delta_{m,2}\right)\\
 & +J_2\bigg((B^2_aB^1_b+B^3_aB^4_b)\delta_{m,1} 
  +\left(B^2_aB^3_b+B^1_aB^4_b\right)\delta_{m,2}\bigg).
  \nonumber
   \end{aligned}
  \end{equation} 
  Here, $a,b,c,d =1,2,3$ and $\alpha,\beta,\gamma=x,y,z$,   
$j=2$ and $1$ for the regions R$_1$ and R$_2$, respectively.    
 
In order to obtain the three branches of triplet dispersions,
$I_BH_{\bold{k}}$ (Eq \ref{hk}) has been diagonalized, where
$I_B$=Diag[$1,1,1,-1,-1,-1$]$_{(6\times 6)}$. 
Characteristic equation of the matrix ${ I_B}H_{\bold{k}}$, and the triplet
dispersions, $\Omega_{b,\bold{k}},\; b=1,2,3$, are given by 
 \begin{equation}
  \begin{aligned}
& \Omega_\bold{k}^6+ a_{2,\bold{k}}\Omega_\bold{k}^4+a_{1,\bold{k}}\Omega_\bold{k}^2+a_{0,\bold{k}}=0, \\
& \Omega_{b,\bold{k}}=\left[2\sqrt{-Q_\bold{k}}\cos{\left(\frac{\theta}{3}-\frac{2\pi(b-1)}{3}\right)}-\frac{a_{2,\bold{k}}}{3}\right]^\frac{1}{2},\;{\rm where} \\
    & Q_\bold{k}=\frac{3a_{1,\bold{k}}-a^2_{1,\bold{k}}}{9},\;\;
    \cos(\theta)=\frac{-R_\bold{k}}{Q_\bold{k}\sqrt{-Q_\bold{k}}},\\
    &R_\bold{k}=\frac{9a_{2,\bold{k}}a_{1,\bold{k}}-27a_{0,\bold{k}}-2a^3_{1,\bold{k}}}{54}.
    \nonumber 
 \end{aligned}
\end{equation}
Expressions of the coefficients, $a_{i,\bold{k}}$, are mentioned below.
 \begin{equation}
  \begin{aligned}
   a_{2,\bold{k}}&=-\left(w^2_{11,\bold{k}}+w^2_{22,\bold{k}}+w^2_{33,\bold{k}}\right), \\
   a_{1,\bold{k}}&=w^2_{11,\bold{k}}w^2_{22,\bold{k}}+w^2_{11,\bold{k}}w^2_{33,\bold{k}}+w^2_{22,\bold{k}}w^2_{33,\bold{k}}\\
& -4(Y^{12}_\bold{k})^2\left(X^{11}_\bold{k}-Y^{11}_\bold{k}\right)\left(X^{22}_\bold{k}-Y^{22}_\bold{k}\right)\\
& -4(Y^{23}_\bold{k})^2\left(X^{22}_\bold{k}-Y^{22}_\bold{k}\right)\left(X^{33}_\bold{k}-Y^{33}_\bold{k}\right)\\
& -4(Y^{13}_\bold{k})^2\left(X^{11}_\bold{k}-Y^{11}_\bold{k}\right)\left(X^{33}_\bold{k}-Y^{33}_\bold{k}\right), \\
  a_{0,\bold{k}}&=\left(X^{11}_\bold{k}-Y^{11}_\bold{k}\right)\left(X^{22}_\bold{k}-Y^{22}_\bold{k}\right)\left(X^{33}_\bold{k}-Y^{33}_\bold{k}\right)\\
&  \big[ 4(Y^{12}_\bold{k})^2\left(X^{33}_\bold{k}+Y^{33}_\bold{k}\right) 
  + 4(Y^{23}_\bold{k})^2\left(X^{11}_\bold{k}+Y^{11}_\bold{k}\right)\\
 & + 4(Y^{13}_\bold{k})^2\left(X^{22}_\bold{k}+Y^{22}_\bold{k}\right)-16Y^{12}_\bold{k}Y^{13}_\bold{k}Y^{23}_\bold{k}\\
 &-\left(X^{11}_\bold{k}+Y^{11}_\bold{k}\right)\left(X^{22}_\bold{k}+Y^{22}_\bold{k}\right)\left(X^{33}_\bold{k}+Y^{33}_\bold{k}\right)\big],\nonumber
   \end{aligned}
\end{equation}
where $w^2_{aa,\bold{k}}=(X^{aa}_\bold{k})^2-(Y^{aa}_\bold{k})^2$ with $a=1,2,3$.

Using the procedure described in \cite{Colpa}, 
the Bogoliubov coefficients has been determined.
The Bogoliubov coefficients are  
$u^{ab}_\bold{k}=(\phi_\bold{k}^{ab}+\psi_\bold{k}^{ab})/2$, and 
  $v^{ab}_\bold{k}=(\phi_\bold{k}^{ab}-\psi_\bold{k}^{ab})/2$,  
with $a,b=1,2,3$, where,
 \begin{equation}
  \begin{aligned}
\phi_\bold{k}^{1a}&=x_{a,\bold{k}}\sqrt{X_\bold{k}^{11}-Y_\bold{k}^{11}},\;
\phi_\bold{k}^{2a}=y_{a,\bold{k}}\sqrt{X_\bold{k}^{22}-Y_\bold{k}^{22}},\\
\phi_\bold{k}^{3a}&= z_{a,\bold{k}} \sqrt{X_\bold{k}^{33}-Y_\bold{k}^{33}},\\
\psi_\bold{k}^{1a}&=\frac{1}{\Omega_{a,\bold{k}}}\bigg(x_{a,\bold{k}}(X_\bold{k}^{11}+Y_\bold{k}^{11})\sqrt{X_\bold{k}^{11}-Y_\bold{k}^{11}}  \\
+ 2&y_{a,\bold{k}}Y_\bold{k}^{12}\sqrt{X_\bold{k}^{22}-Y_\bold{k}^{22}} 
+2z_{a,\bold{k}} Y_\bold{k}^{13}\sqrt{X_\bold{k}^{33}-Y_\bold{k}^{33}}\bigg),\\
\psi_\bold{k}^{2a}\!\!&=\!\frac{1}{\Omega_{a,\bold{k}}}\!\bigg(2 x_{a,\bold{k}} Y_\bold{k}^{12}\sqrt{X_\bold{k}^{11}-Y_\bold{k}^{11}}\\
\!+\!&y_{a,\bold{k}} (X_\bold{k}^{22}+Y_\bold{k}^{22})\sqrt{X_\bold{k}^{22}-Y_\bold{k}^{22}} 
+2z_{a,\bold{k}} Y_\bold{k}^{23}\sqrt{X_\bold{k}^{33}-Y_\bold{k}^{33}}\bigg),\\
\psi_\bold{k}^{3a}&=\frac{1}{\Omega_{a,\bold{k}}}\bigg(2 x_{a,\bold{k}} Y_\bold{k}^{13}\sqrt{X_\bold{k}^{11}-Y_\bold{k}^{11}}
+2y_{a,\bold{k}} Y_\bold{k}^{23}\sqrt{X_\bold{k}^{22}-Y_\bold{k}^{22}} \\
&+z_{a,\bold{k}} (X_\bold{k}^{33}+Y_\bold{k}^{33})\sqrt{X_\bold{k}^{33}-Y_\bold{k}^{33}}\bigg),\\
x_{a,\bold{k}}&=\frac{M_{a,\bold{k}}}{\sqrt{G_{a,\bold{k}}}},\;
y_{a,\bold{k}}=\frac{1}{\sqrt{G_{a,\bold{k}}}},\;
z_{a,\bold{k}}=\frac{N_{a,\bold{k}}}{\sqrt{G_{a,\bold{k}}}},\\
M_{a,\bold{k}}&=\frac{A_{\bold{k}}C_{\bold{k}}-(w^2_{22,\bold{k}}-\Omega^2_{a,\bold{k}})B_{\bold{k}}}
{A_{\bold{k}}B_{\bold{k}}-(w^2_{11,\bold{k}}-\Omega^2_{a,\bold{k}})C_{\bold{k}}}, \\
N_{a,\bold{k}}&=\frac{C_{\bold{k}}+B_{\bold{k}}M_{a,\bold{k}}}{(\Omega^2_{a,\bold{k}}-w^2_{33,\bold{k}})},\;
G_{a,\bold{k}}=\Omega_{a,\bold{k}}\left[1+M^2_{a,\bold{k}}+N^2_{a,\bold{k}}\right],\\
A_{\bold{k}}&=2 Y_\bold{k}^{12}\sqrt{X_\bold{k}^{11}-Y_\bold{k}^{11}}\sqrt{X_\bold{k}^{22}-Y_\bold{k}^{22}},  \\
B_{\bold{k}}&=2 Y_\bold{k}^{13}\sqrt{X_\bold{k}^{11}-Y_\bold{k}^{11}}\sqrt{X_\bold{k}^{33}-Y_\bold{k}^{33}}, \\
C_{\bold{k}}&=2 Y_\bold{k}^{23}\sqrt{X_\bold{k}^{22}-Y_\bold{k}^{22}}\sqrt{X_\bold{k}^{33}-Y_\bold{k}^{33}}. \nonumber
    \end{aligned}
 \end{equation}

  \end{document}